\documentclass[aps,prl,reprint,superscriptaddress,longbibliography]{revtex4-2}

\usepackage{graphicx}
\usepackage{gensymb} 
\usepackage{xcolor}

\begin{document}

\title{Campbell penetration depth in a single crystal of heavy fermion superconductor CeCoIn$_5$}

\author{Hyunsoo Kim}
\affiliation{Ames National Laboratory, Ames 50011, IA, USA}
\affiliation{Department of Physics \& Astronomy, Iowa State University, Ames 50011, USA}
\affiliation{Department of Physics, Missouri University of Science and Technology, Rolla, MO 65409, USA}

\author{Makariy A. Tanatar}
\affiliation{Ames National Laboratory, Ames 50011, IA, USA}
\affiliation{Department of Physics \& Astronomy, Iowa State University, Ames 50011, USA}

\author{Cedomir~Petrovic}
\affiliation{Condensed Matter Physics and Materials Science Department, Brookhaven National Laboratory, Upton, New York 11973, USA}
\affiliation{Shanghai Key Laboratory of Material Frontiers Research in Extreme Environments (MFree), Shanghai Advanced Research in Physical Sciences (SHARPS), Pudong, Shanghai 201203, China}
\affiliation{Department of Nuclear and Plasma Physics, Vinca Institute of Nuclear Sciences, University of Belgrade, Belgrade 11001, Serbia}

\author{Ruslan Prozorov}
\affiliation{Ames National Laboratory, Ames 50011, IA, USA}
\affiliation{Department of Physics \& Astronomy, Iowa State University, Ames 50011, USA}

\date{24 April 2026} 

\begin{abstract}
The temperature and magnetic field dependent magnetic penetration depth, $\lambda_m(T,H)$, was measured in a single crystal of a heavy fermion superconductor CeCoIn$_5$ using a frequency-domain tunnel diode resonator. 
In addition to the London penetration depth, which yields the superfluid density, measurements in a finite DC magnetic field provide Campbell penetration depth, $\lambda_C(T,H)$, which is directly linked to the true (unrelaxed) critical current density, $J_c$. The measured $\lambda_C(H)$ in CeCoIn$_5$ deviates significantly from the conventional $\sim \sqrt{H}$ behavior, and its slope changes abruptly at the characteristic magnetic field values. Considering that our sample is in the clean limit, we interpret this deviation as a fingerprint of the vortex lattice symmetry change. The temperature dependence $J_c(T)$ of CeCoIn$_5$ calculated from $\lambda_C(T)$ is nearly $T$-linear over the entire temperature range, also in stark contrast to expectations in a conventional type-II superconductor. Our results provide new evidence for unconventional superconductivity in CeCoIn$_5$ from the never-before-measured Campbell penetration depth. 
\end{abstract}

\maketitle

\section{\label{sec:intro}Introduction}

The superconductivity of the heavy fermion CeCoIn$_5$ has a number of unusual features, keeping this material in the focus of active interest even 25 years after its discovery \cite{Pfleiderer2009}. CeCoIn$_5$ is a member of the CeTIn$_5$ family (T=Co,Rh,Ir), commonly referred to as ``115" compounds \cite{Hegger2000,Petrovic2001CeIrIn5,Petrovic2001CeCoIn5,Thompson2012}. The parent compound, CeRhIn$_5$, orders magnetically below $T_N~\approx 3.8$~K, and superconductivity with $T_c$ up to 2.5~K is induced by applying a modest pressure of approximately 2 GPa \cite{Hegger2000}. The superconductivity in CeCoIn$_5$ is observed at ambient pressure, but many studies show that it is also in proximity to a magnetic order \cite{Young2007, Gofryk2012, Stock2018}. This proximity indicates that superconductivity may be magnetically mediated, as suggested for many heavy fermion compounds \cite{Mathur1998}. 
Indeed, a number of experimental studies have shown superconducting properties consistent with the presence of line nodes in the superconducting energy gap \cite{Kohori2001, Movshovich2001, Ozcan2003, Rourke2005, Kim2015}, including the momentum dependence of quasiparticle interference \cite{Allan2013,Zhou2013} and angular variation of both thermal conductivity measurements \cite{Izawa2001,Vorontsov2006} and heat capacity measurements \cite{Aoki2004,Vorontsov2006} in a magnetic field.

Magnetism in CeCoIn$_5$ may be induced by an application of a magnetic field \cite{Young2007} and by hole-doping with the substitution of In with Cd and Hg \cite{Gofryk2012,Stock2018}. The presence of spin resonance of commensurate antiferromagnetic fluctuations with $\mathbf{Q_0}=(1/2,1/2,1/2)$ is indicative of strong coupling between $f$-electron magnetism and superconductivity, with magnetic ordering induced at the same wavevector \cite{Stock2008, Stock2018}. 
This is consistent with a $d$-wave order parameter satisfying $\Delta(\mathbf{q}+\mathbf{Q_0})=-\Delta(\mathbf{q})$ \cite{Stock2008, Gu2019, Chubukov2008}.
Early measurements of London penetration depth are also consistent with the $d$-wave pairing, while there have been reports that deviate from a simple $d$-wave scenario \cite{Kogan2009,Howald2013}.
In particular, Yb-doping induces a power-law behavior which cannot be reconciled with a symmetry-imposed line nodes \cite{Kim2015}. 
Therefore, a novel approach for probing the superconducting pairing symmetry in CeCoIn$_5$ would help to clarify the physics of these fascinating materials.

The symmetry of the vortex lattice (VL) is linked to the intrinsic electronic anisotropy of the system studied and, in some cases, to the structure of the superconducting gap \cite{Amin1998,Ichioka1999}. In most superconductors, the triangular vortex lattice has the lowest ground state energy \cite{Abrikosov2017,Cribier1964,Essmann1967}, which is generally robust \cite{Hoffmann2022}, particularly in a fully-gapped superconductor such as LiFeAs \cite{Kim2011,Hoffmann2022}.
However, distortions of the simple lattice structure were observed in many cases \cite{Das2012MgB2,Hirano2013,Leishman2021,Agterberg1998PRL,Agterberg1998PRB,Riseman1998} and its microscopic origins were attributed to anisotropy of the Fermi surface \cite{Das2012MgB2,Hirano2013,Leishman2021} and anisotropy of the superconducting gap \cite{Agterberg1998PRL,Agterberg1998PRB,Riseman1998}. 
When the VL is tuned by temperature and field, its transformation was observed in various unconventional superconductors \cite{Huxley2000,Brown2004,Bianchi2008,Avers2020,Avers2022}, forming a rich VL $H$-$T$ phase diagram \cite{Huxley2000,Brown2004,Bianchi2008,Avers2020,Avers2022,Hoffmann2022}.

The vortex state in CeCoIn$_5$ has been extensively studied with a particular focus on the effects of the quantum critical point (QCP) and the relationship between magnetism and superconductivity \cite{Hu2012, Jang2016, Wulferding2020}. 
In a transport study, the vortex core resistivity was found to increase with decreasing temperature in the mixed state, signaling suppressed the AFM order within the vortex core \cite{Hu2012}. 
The bulk pinning effect is consistent with an asymmetric peak effect, which cannot be explained by the pinning of the vortex due to defects, and the asymmetric peak effect cannot be reconciled with conventional scenarios for the peak effect \cite{Jang2016}. 
In addition, remarkable field evolution of the VL symmetry was found in multiple studies \cite{Eskildsen2003, Bianchi2008, Ohira-Kawamura2008, White2010}, which was attributed to strong Pauli paramagnetism \cite{Ohira-Kawamura2008}, strongly field-dependent coherence length \cite{DeBeer-Schmitt2006}, and the $d$-wave order parameter \cite{Eskildsen2003}. 
Furthermore, evidence for nodal superconductivity was also found in the field orientation dependence of VL \cite{Das2012CeCoIn5}.
Therefore, studying the VL can be a valuable contribution to understanding the superconducting gap structure.

\begin{figure*}[tb]
\includegraphics[width=1.0\linewidth]{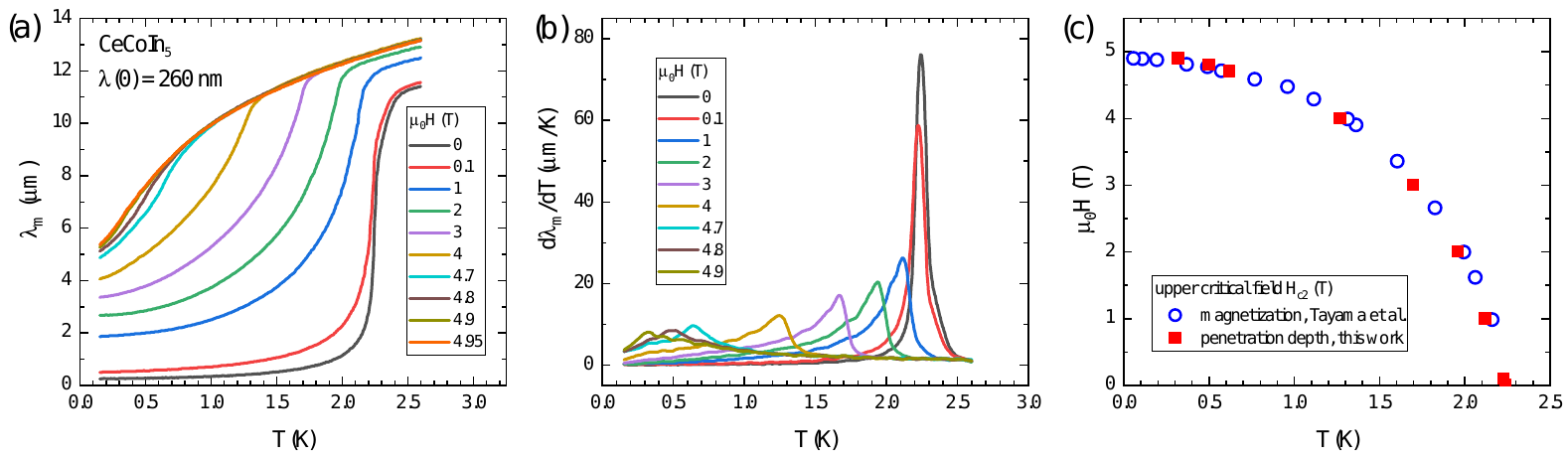}%
\caption{
\label{fig1} 
(a) Temperature-dependent in-plane magnetic penetration depth $\lambda_m(T)$ in a single crystal CeCoIn$_5$ sample with $H_{ac}$ and $H_{dc}$ applied along the crystallographic $c$-axis. Different curves correspond to different values of the applied dc magnetic field indicated in the legend. (b) First temperature-derivative $d\lambda_m/dT$ of $\lambda_m(T)$ shown in panel (a). The sharp peaks correspond to superconducting phase transitions for each applied field. 
(c) The superconducting $H$-$T$ phase diagram. The derivative maximum criteria was compared to bulk measurements determined at a discontinuous jump in dc magnetization measurements by T. Tayama {\it et al.} \cite{Tayama2002}.
}
\end{figure*}

The small-angle neutron scattering (SANS) and scanning tunnel microscopy (STM) techniques have been commonly employed to study the transformation of VL \cite{Avers2022,Hoffmann2022}, but their sample requirements limit the wider application of these techniques. In particular, SANS usually requires very large samples, and STM requires cleavable crystals. Conventional magnetization and ac susceptibility are sensitive to both vortex pinning and vortex relaxation, making it difficult to interpret the results if vortex lattice transformations are involved.

The Campbell penetration depth, on the other hand, has been successfully employed to study the mixed state \cite{Prommapan2011,Kim2013,Kim2021}. In the Campbell regime, the ac magnetic field must have a small enough amplitude so that the ac field does not push vortices out of the pinning wells. In this ``linear ac response" regime, the penetration depth is governed by the elastic properties of the Abrikosov vortex lattice, where Hook's law is applicable. In the presence of a dc magnetic field that creates a vortex lattice, the ac perturbation due to a superimposed ac magnetic field is attenuated exponentially from sample edge into its interior, $B_{ac}\left(x\right)\propto\mu H_{ac}e^{-x/\lambda_{C}}$, with a characteristic depth $\lambda_{C}$. Taking into account the always present London penetration depth, $\lambda_L$, the effective measured penetration depth is $\lambda_m^2=\lambda_L^2+\lambda_C^2$. Conveniently, $\lambda_L=\lambda_m|_{H=0}$ is obtained from the same measurement performed first without any applied dc magnetic field \cite{Campbell1969,Campbell1971,Brandt1995,Prozorov2003e,Willa2015prb,Willa2015,Willa2016}. This looks simple, but it is very difficult to measure $\lambda_{C}$, due to the rigorous constraint on the ac field amplitude on one hand, and sensitivity on the other. Commonly measured amplitude-domain ac susceptibility is usually in the regime where vortices are displaced at large distances \cite{Gomory1997}, and their collective response is described by the critical-state model, such as the Bean model \cite{Bean1962, Bean1964, Kim2021}.

If the Campbell penetration depth is obtained, it can be used to determine the ``true critical current", $J_c$,  as opposed to the usually measured ``persistent current", $J_p$, which is always much smaller due to exponentially fast magnetic relaxation at the initial stages of vortex motion. This is because $J_c$ is only a parameter characterizing the pinning potential and is best estimated in the situation when the vortex distribution is uniform, after cooling in a magnetic field. Then, $\lambda_{C}^{2}=Hr_{p}/J_{c}$. Here $r_{p}$ is the radius of the pinning potential, usually assumed to be of the order of the coherence length, and $H$ is the applied external dc magnetic field. 
To discuss the superconducting gap structure, one would need the ``deparing current density", $J_d$, determined by the critical velocity through the Landau criterion, giving direct access to the superfluid density $n_s\propto J_d$ \cite{Bardeen1958,Gorkov1959,Tinkham2004}. However, $J_d$ cannot be directly measured, and therefore we must rely on the ``true critical current" and its connection to the gap structure through properties of the vortex lattice.

In this work, we use a precision tunnel-diode resonator-based (TDR) rf-technique to study the temperature and field evolution of the rf penetration depth, $\lambda_m(T,H)$, calculate Campbell penetration depth, $\lambda_C(T,H)$, and use it to calculate the theoretical critical current density. We detected special characteristic points in $\lambda_C(T,H)$, which allow us to construct the $H$-$T$ phase diagram of CeCoIn$_5$. This approach was successfully applied to various superconductors \cite{Kim2018rsi,Prommapan2011,Kim2013,Kim2021}. Our results are consistent with the previous SANS experiment \cite{Bianchi2008} and strongly support unconventional superconductivity in the heavy fermion CeCoIn$_5$ superconductor.


\section{\label{sec:exp}Experiment}

Single crystals of CeCoIn$_5$ were synthesized from indium flux by combining stoichiometric amounts of Ce and Co with excess indium in an alumina crucible and encapsulating the crucible in an evacuated quartz ampule heated to 1150\degree C. 
After homogenizing the molten material, the ampule is cooled to 450\degree C, at which temperature the excess In flux is decanted with a centrifuge. 
The resultant single crystals are well-separated, faceted platelets with characteristic dimensions 3 mm $\times$ 3 mm $\times$ 0.1 mm \cite{Petrovic2001CeCoIn5}.

The variation in the magnetic penetration depth of the radio-frequency (rf) with respect to its value at the minimum temperature, $T_{min}\approx 150\;\text{mK}$, $\Delta\lambda_{m}$, was measured in a dilution refrigerator using a tunnel diode resonator (TDR) technique \cite{Degrift1975, Kim2018rsi,Prozorov2006, Prozorov2011}.
The key component of the technique is an $LC$-tank circuit powered by a tunnel diode. The effective inductance $L_\textmd{\tiny eff}$ of the circuit is a combination of contributions from the inductor coil itself and the sample. Variation of sample ac magnetic susceptibility (including ac screening due to skin effect in the normal state) with temperature and magnetic field leads to changes in the magnetic inductance, $\Delta L_\textmd{\tiny eff}$, and hence to the shift of the resonant frequency, $\Delta f$. 
For a coil with volume $V_c$, a sample with magnetic susceptibility $\chi_m$ and volume $V_s$, the shift is~\cite{Vannette2008}
\begin{equation}
\frac{\Delta f}{f_0}\approx -\frac{1}{2}\frac{V_s}{V_c}4\pi\chi_m.
\end{equation}
For an infinite slab of thickness $2d$ in a parallel field,
\begin{equation}
\frac{\Delta f}{f_0}\approx \frac{1}{2} \frac{V_s}{V_c} \left[1-\textmd{Re}\left\{ \frac{\tanh (\alpha d)}{\alpha d}\right\} \right]
\end{equation}
\noindent where $\alpha=(1-i)/\delta$ \cite{Hardy1993,Kim2018rsi}.

The sample with dimensions (0.5 mm $\times$ 0.8 mm $\times$ 0.1 mm) positioned with the shortest direction being $c$-axis along $H_{ac}$ was mounted on a sapphire rod and inserted into a 2 mm inner diameter copper coil that (when empty) produces rf excitation field with amplitude $H_{ac}\approx 10$ mOe and frequency of $f_{0}\approx20$ MHz. The shift of the resonant frequency (in cgs units), $\Delta f(T)=-G4\pi\chi(T)$, where $\chi(T)$ is the differential magnetic susceptibility, $G\approx f_{0}V_{s}/2V_{c}(1-N)$ is a constant, $N$ is the effective demagnetization factor, $V_{s}$ is the sample volume and $V_{c}$ is the coil volume \cite{Prozorov2006}. 
The constant $G$ was determined from the frequency change by physically pulling the sample out of the coil. With the characteristic sample size, $R$, $4\pi\chi=(\lambda_{m}/R)\tanh(R/\lambda_{m})-1$, from which $\Delta\lambda_{m}$ can be obtained \cite{Prozorov2006}.

\section{\label{sec:rs}Results and Discussion}

\subsection{Magnetic penetration depth and $H$-$T$ phase diagram}

Figure 1(a) shows the magnetic field evolution of $\lambda_m(T)$ from $\mu_0 H_{dc}=0$ to 5 T. 
In the absence of a magnetic field, the single crystal sample exhibits a step-like superconducting phase transition near $T=2.3$ K where the London penetration depth, $\lambda_L$, diverges.
The temperature dependence of $\lambda_L$ in the Meissner state of unconventional superconductors is determined by the symmetry of the superconducting gap. 
In particular, a power-law behavior $\Delta\lambda(T)\sim T^n$ best describes its low-temperature behavior below typically $0.3T_c$.
For a line-nodal superconductor, the exponent $n$ varies between 1 and 2 depending on the impurity scattering \cite{Hirschfeld1993} and the nature of the scattering center \cite{Cho2022}.
The low-temperature behavior of $\lambda_L$ in our samples can be found elsewhere \cite{Kim2015} and is consistent with the existence of lines of nodes in the gap \cite{Ozcan2003, Ormeno2002, Truncik2013, Hashimoto2013, Kim2015}.
The normal state response, above $T_c$, is determined by the skin depth,
\begin{equation}
    \delta (T) = \sqrt{\frac{\rho(T)}{\pi f \mu}},
\end{equation}
where $\mu$ is the magnetic permeability of the sample. 
Here $\rho(T)$ is electrical resistivity, and $\rho(2.5 \textmd{ K})$ ranges from 3.0 to 3.5 $\mu\Omega$ cm \cite{Petrovic2001CeCoIn5, Paglione2003}.

In a finite magnetic field, the superconducting transition temperature $T_c(H)$ is lowered due to the orbital pair-breaking effect, and the superconducting transition becomes progressively broader with the increasing field.
Whereas a sharp first-order phase transition is expected near the Pauli limiting upper critical field $H_{c2}(0)=4.95$ T, we did not observe such a sharp feature.
We attribute this to the clean nature of the sample where the normal skin depth, $\delta(T)$, is much smaller than the variation of the penetration depth near $H_{c2}(T)$ at this field range. 
Similar broad transitions in a magnetic field were observed in a stoichiometric Fe-based superconductor KFe$_2$As$_2$ \cite{Kim2018rsi}.
For $\mu_0 H\gtrsim 4.95$ T, the temperature variation of $\Delta f$ in CeCoIn$_5$ is determined by the normal skin effect. 
The noticeable sub-linear behavior of $\delta(T)$ in temperature is consistent with the non-Fermi liquid normal state due to an antiferromagnetic quantum critical point near $H_{c2}(0)$ \cite{Paglione2003}.

The broad transitions for measurements in applied magnetic fields, particularly near $H_{c2}(0)$, make the consistent determination of $T_c(H)$ difficult.
To construct the superconducting $H$-$T$ phase diagram, we used the derivative $d\lambda_m/dT$ maximum criterion to determine $T_c(H)$ in a field where it exhibits a well-defined maximum even for $\mu_0H=4.9$ T as shown in Fig. \ref{fig1}(b).
This criterion is justified because of the diverging nature of the penetration depth at the transition.
The same criterion was used previously in various superconductors \cite{Kim2018rsi, Kim2021} to construct $H$-$T$ phase diagrams in KFe$_2$As$_2$ \cite{Kim2018rsi} and YPtBi \cite{Kim2021} that were compatible with the bulk magnetization measurements and zero resistivity criterion. 
In Fig.~\ref{fig1}(c), the superconducting $H$-$T$ phase diagram of CeCoIn$_5$ from the penetration depth measurements is compared to that based on the dc magnetization measurements taken from Ref. \cite{Tayama2002}. 
It shows excellent agreement, and therefore $T_c(H)$ determined at the $d\lambda_m/dT$ maximum represents the bulk superconducting transition.

\subsection{Campbell penetration depth}

\begin{figure}
\includegraphics[width=0.9\linewidth]{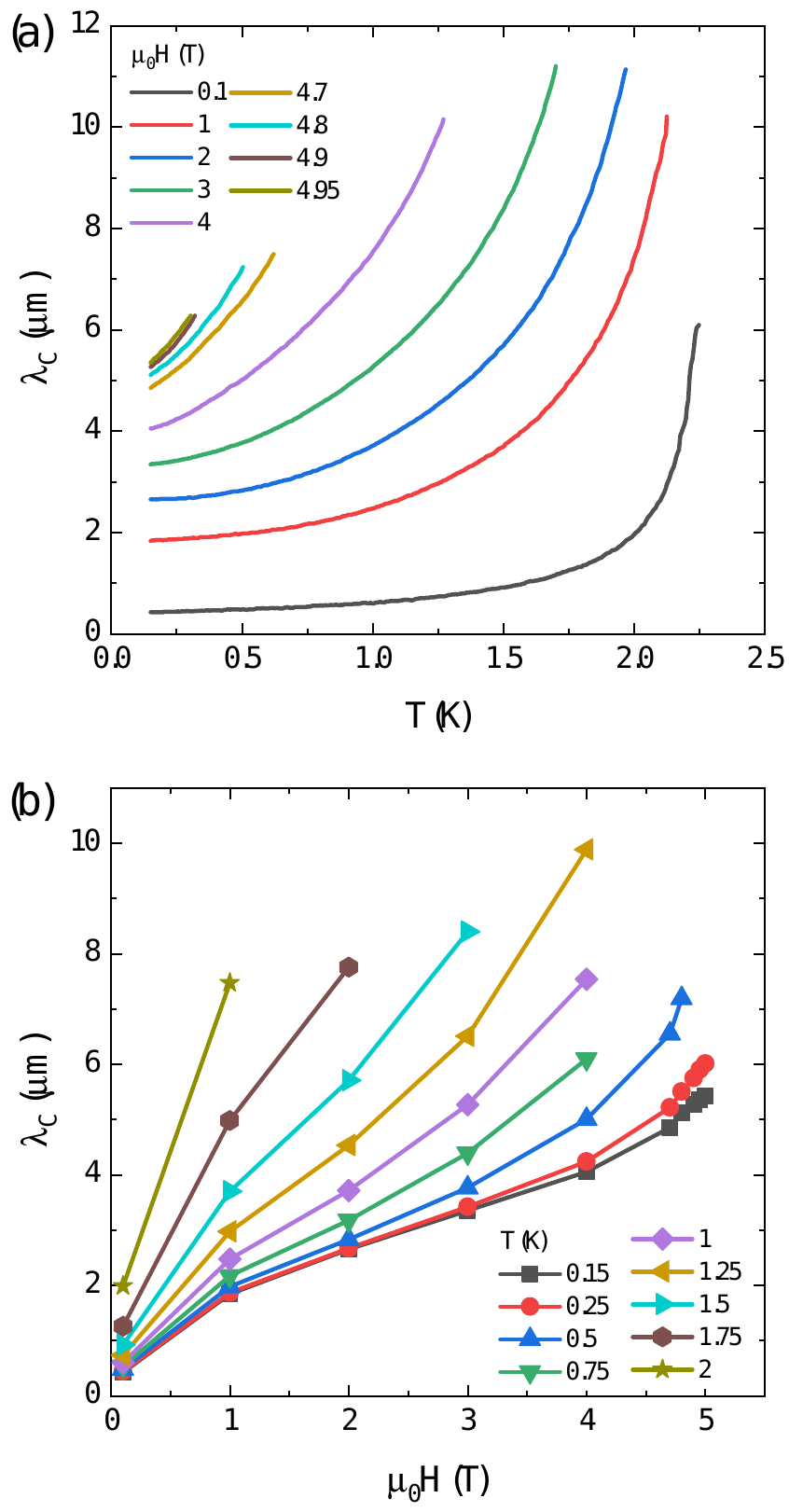}%
\caption{\label{fig2} 
(a) Temperature-variation of Campbell penetration depth $\lambda_C$ at various fixed fields in CeCoIn$_5$. (b) Isothermal field-variation of $\lambda_C$ at various fixed temperatures.
}
\end{figure}

When a magnetic field greater than the lower critical field $H_{c1}$ is applied, a vortex lattice (VL) forms. In this case, the measured $\lambda_m$ consists of the London penetration depth $\lambda_L$ and Campbell penetration depth $\lambda_C$ as detailed in the introduction section.
We determined the Campbell penetration depth by using a relation, $\lambda^2_C(T)=\lambda^2_m(T)-\lambda^2_L(T)$ in the approximation of a linear elastic response of a VL to a small amplitude $H_{ac}$ \cite{Campbell1969, Campbell1971, Brandt1991, Brandt1995}. 
$\lambda^2_L(T)$ corresponds to the zero-field $\lambda^2_m(T)$.
Here we used a relation, $\lambda_L(T)=\lambda_L(0) + \Delta \lambda_L(T)$ with $\lambda_L(0)=260$ nm that is an average value of reported values \cite{Ozcan2003, Ormeno2002, Chia2003, Howald2013, Wulferding2020}.
The calculated $\lambda_C(T)$ curves in CeCoIn$_5$ at various fixed fields are shown in Fig. \ref{fig2}(a). 
The $\lambda_C(T)$ exhibits a monotonic increase as the temperature is raised in all applied magnetic fields, signaling the increasing propagation of the VL modulation with temperature at a fixed field due to softening of the vortex lattice. 
Near $T_c(H)$, $\lambda_C$ exhibits a rapid rise due to the diverging nature of the measured penetration depth.

To investigate the elasticity of VL in CeCoIn$_5$, we used the single-vortex relation, 
\begin{equation}\label{eq:Campbell}
\lambda^2_C=\frac{\phi_0 H}{\alpha},
\end{equation}
obtained assuming the quadratic pinning potential 
\begin{equation}
    U(u)=\frac{1}{2}\alpha u^2.
\end{equation}
Here $\phi_0$ is the magnetic flux quantum and $\alpha$ is the Labusch parameter \cite{Labusch1969, Brandt1991, Brandt1995}, representing the pinning strength, and $u$ is the displacement of the vortex from the pinning center \cite{Kim2021}. 
Isothermal $\lambda_C(H)$ curves at various fixed temperatures are shown in Fig. \ref{fig2}(b).
The monotonic increase of $\lambda_C$ with field is consistent with the VL becoming softer and therefore more susceptible to the ac magnetic field.

\begin{figure}
\includegraphics[width=0.9\linewidth]{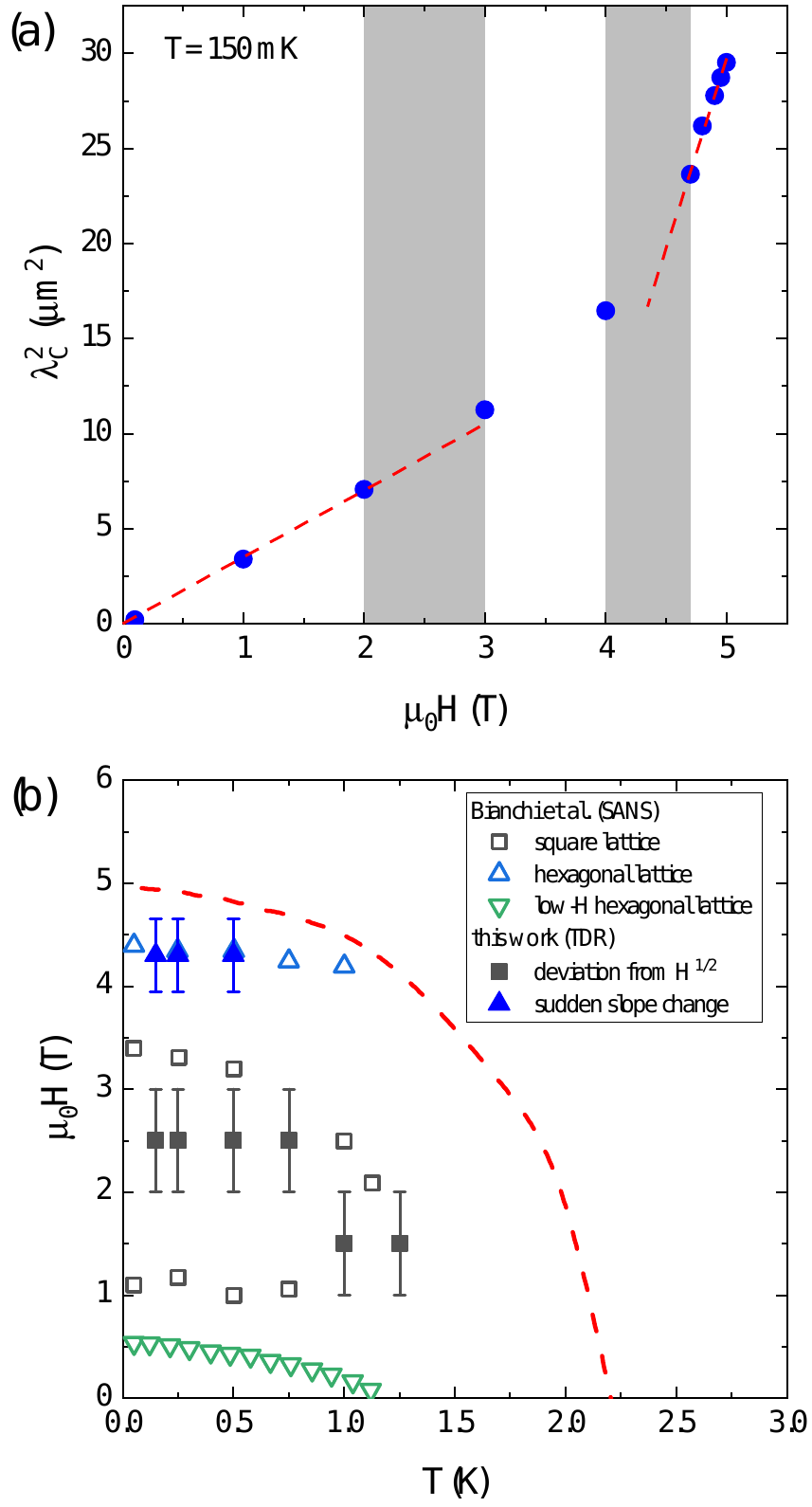}%
\caption{\label{fig3} (a) Field-variation of $\lambda^2_C$ in CeCoIn$_5$ at $T=150$ mK  (blue dots). The dashed straight lines are depicted for the eyes. The vertical shaded regions indicate possible deviation of $\Delta\lambda_C(H)$ from $\sqrt{H}$ variation. (b) Vortex line lattice $H$-$T$ phase diagram in CeCoIn$_5$. The vortex phases, indicated with the open symbols, were taken from Ref. \cite{Bianchi2008}, and the closed symbols were determined from the observed anomalies in $\lambda_C(H)$ in this work.
 }
\end{figure}

According to Eq.~(\ref{eq:Campbell}), the Campbell penetration depth varies as $\lambda_C\sim \sqrt{H}$ in the single-vortex pinning regime (field-independent Labusch constant). 
To verify the relationship, we plot $\lambda^2_C$ vs. $H$ at $T=150$ mK in Fig. \ref{fig3}(a). In the low field region, up to 2 T, it is consistent with the $\sqrt{H}$-variation. However, a noticeable deviation is observed between 2 and 3 T, and a more rapid slope-change was observed between 4 and 5 T. The anomalous field ranges are indicated by the gray vertical bars in Fig. \ref{fig3}(a). 
We determined the anomalous magnetic field dependence in other $\lambda_C$-isotherms and summarized the result in Fig. \ref{fig3}(b).
The anomalous field points, represented by closed squares and triangles, are laid over a known VL phase diagram, constructed with small-angle neutron scattering measurements \cite{Bianchi2008}. 
Various magnetic field-induced VL phases are indicated by open symbols (see the legend in Fig. \ref{fig3}). The anomalous behavior detected by the $\lambda_C$ measurements shows reasonable agreement with the neutron scattering experiment, indicating the observed deviation from $\lambda_C \sim \sqrt{H}$ is related to VL transformations near QCP.

\subsection{Theoretical critical current density}

\begin{figure}
\includegraphics[width=1.0\linewidth]{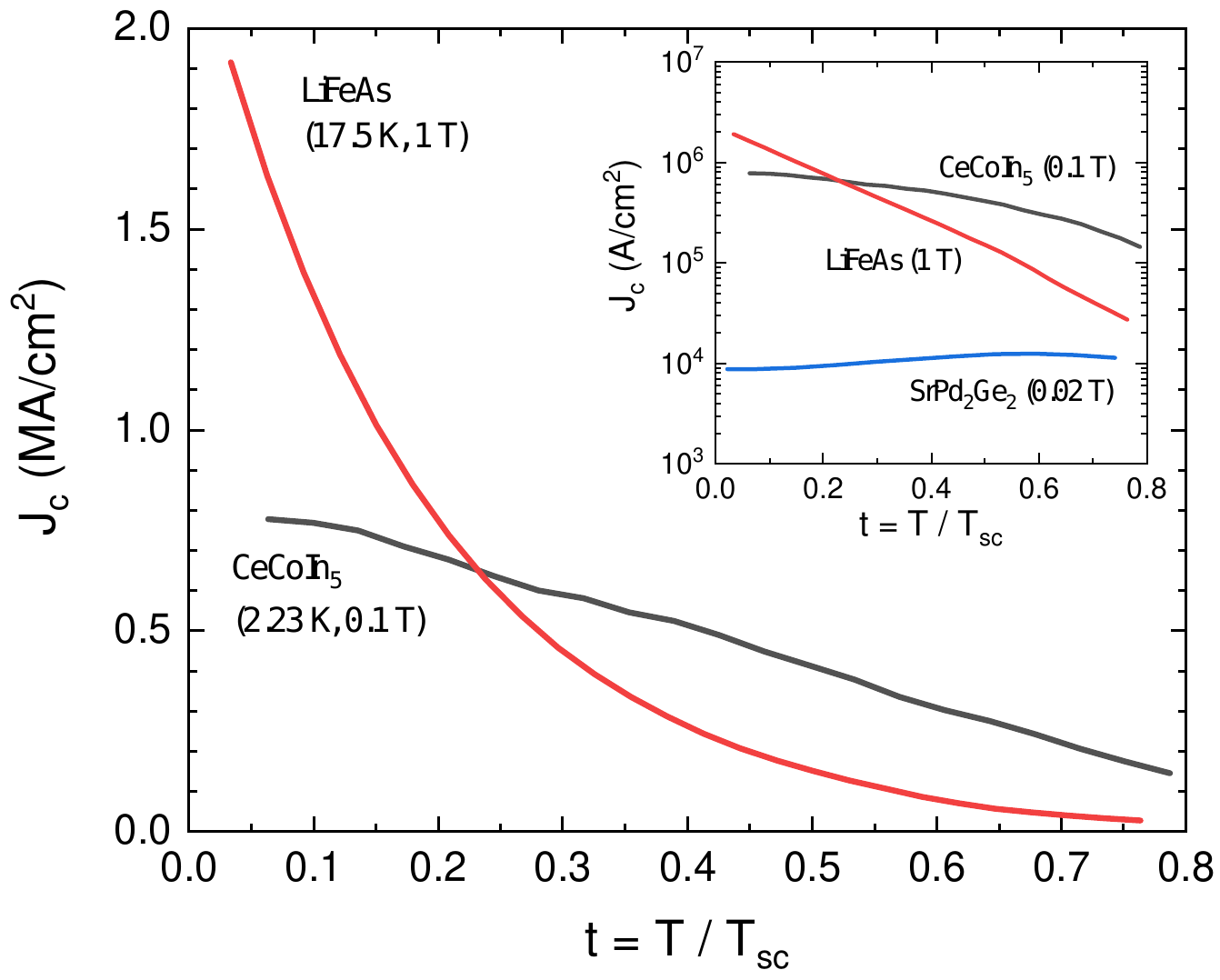}%
\caption{\label{fig4} Temperature-dependent theoretical critical current density $J_c(t)$ with a dc magnetic field $\mu_0 H_0 =$ 0.1 T in CeCoIn$_5$ where $t$ is the reduced temperature $T/T_c$. $J_c(T)$ in LiFeAs and SrPd$_2$Ge$_2$ are shown for comparison, which are taken in $\mu_0 H_0 =$ 1 T and 0.02 T, respectively. The magnetic fields for each superconductor are the minimum applied field for the measurement of $J_c$.
 }
\end{figure}

In a single-vortex pinning model, the equilibrium displacement of the vortex from the pinning center is determined where the pinning force is balanced with the Lorentz force, at which the theoretical critical current density $J_c$ is defined.
Knowledge of $\lambda_C$ allows the determination of $J_c$ in the strong pinning limit by using the following relation, 
\begin{equation}
J_c = \frac{H r_p}{\lambda_C^2} 
\end{equation}
where $r_p$ represents the radius of the vortex pinning potential \cite{Campbell1969, Campbell1971, Kim2013}. 
The coherence length $\xi$ is often adopted as $r_p$ \cite{Kim2013,Kim2021}. 
We use $\xi(0)=24$ nm \cite{Movshovich2001} and its temperature-dependence is obtained from the Ginzburg-Landau coherence length for a clean superconductor \cite{Tinkham2004}, 
\begin{equation}
    \xi(t)=0.74\frac{\xi(0)}{\sqrt{1-t}}
\end{equation}
where $t= T/T_c$. 
By using the relations above, temperature-dependent $J_c$ can be readily obtained. 

The calculated $J_c$ is shown in Fig. \ref{fig4} as a function of the reduced temperature $t$ for various superconductors. In the main panel, $J_c$ in CeCoIn$_5$ with 0.1 T and $T_c(H)=$ 2.23 K is compared to that in LiFeAs, a fully gapped superconductor with $T_c=18$ K \cite{Kim2011}. 
$J_c$ in LiFeAs was taken with $\mu_0 H=1$ T that reduces the superconducting transition to $T_c(H)\approx 0.97T_c$ where $T_c=18$ K \cite{Prommapan2011}. 
The observed $J_c$ in CeCoIn$_5$ is comparable to that in LiFeAs, which is surprising because $T_c$ of CeCoIn$_5$ is only $0.13T_c$ of LiFeAs. Compared to SrPd$_2$Ge$_2$ with $T_c=2.7$ K \cite{Kim2013}, CeCoIn$_5$ exhibits two orders of magnitude greater $J_c$ as shown in the inset. All curves were obtained by using the same TDR rf-technique with similar frequencies.
We note, for comparison, that the critical current density from magnetic force microscopy measurements at 500 mK,  $9\times 10^4$ A/cm$^2$, is about one order of magnitude smaller \cite{Wulferding2020}. 

\section{\label{sec:dis} Conclusions}

In conclusion, we investigated the magnetic penetration depth $\lambda_m$ in a nodal-superconductor CeCoIn$_5$ using a tunnel diode resonator technique operating at 20 MHz. Both ac and dc magnetic fields were applied along the crystallographic $c$-axis of a single crystalline sample. The temperature-dependence of $\lambda_m(T)$ from 0.15 K to 2.5 K was determined at various fixed magnetic fields up to 5 T, from which the temperature-and field-dependent Campbell penetration depth $\lambda_C(T,H)$ was deduced. While $\lambda^2_C(H)$ varies as $H$-linear in the mixed state of a superconductor, CeCoIn$_5$ exhibits an anomalous behavior where $\lambda^2_C(H)$ deviates from the $H$-linear behavior, and the deviation is associated with changes in symmetry in the vortex line lattice in small angle neutron scattering experiments, which makes the measurement of rf Campbell penetration depth a promising tool for investigation of the vortex line lattice. Finally, we calculate the theoretical critical current density of CeCoIn$_5$ from $\lambda_C$ and compare it with that in LiFeAs and SrPd$_2$Ge$_2$, which are known fully-gapped superconductors.

\section{acknowledgments} 
We thank V.~G.~Kogan for multiple useful discussions. Work in Ames was supported by the U.S. Department of Energy (DOE), Office of Science, Basic Energy Sciences, Materials Science and Engineering Division. Ames Laboratory is operated for the U.S. DOE by Iowa State University under the contract DE-AC02-07CH11358.
Work at BNL (materials synthesis) was supported by the U.S. Department of Energy, Basic Energy Sciences, Division of Materials Science and Engineering, under Contract No. DE-SC0012704. C.P. acknowledges support from the Shanghai Key Laboratory of Material Frontiers Research in
Extreme Environments, China (No. 22dz2260800) and Shanghai Science and Technology Committee, China (No. 22JC1410300).


\begin{thebibliography}{82}%
\makeatletter
\providecommand \@ifxundefined [1]{%
 \@ifx{#1\undefined}
}%
\providecommand \@ifnum [1]{%
 \ifnum #1\expandafter \@firstoftwo
 \else \expandafter \@secondoftwo
 \fi
}%
\providecommand \@ifx [1]{%
 \ifx #1\expandafter \@firstoftwo
 \else \expandafter \@secondoftwo
 \fi
}%
\providecommand \natexlab [1]{#1}%
\providecommand \enquote  [1]{``#1''}%
\providecommand \bibnamefont  [1]{#1}%
\providecommand \bibfnamefont [1]{#1}%
\providecommand \citenamefont [1]{#1}%
\providecommand \href@noop [0]{\@secondoftwo}%
\providecommand \href [0]{\begingroup \@sanitize@url \@href}%
\providecommand \@href[1]{\@@startlink{#1}\@@href}%
\providecommand \@@href[1]{\endgroup#1\@@endlink}%
\providecommand \@sanitize@url [0]{\catcode `\\12\catcode `\$12\catcode
  `\&12\catcode `\#12\catcode `\^12\catcode `\_12\catcode `\%12\relax}%
\providecommand \@@startlink[1]{}%
\providecommand \@@endlink[0]{}%
\providecommand \url  [0]{\begingroup\@sanitize@url \@url }%
\providecommand \@url [1]{\endgroup\@href {#1}{\urlprefix }}%
\providecommand \urlprefix  [0]{URL }%
\providecommand \Eprint [0]{\href }%
\providecommand \doibase [0]{https://doi.org/}%
\providecommand \selectlanguage [0]{\@gobble}%
\providecommand \bibinfo  [0]{\@secondoftwo}%
\providecommand \bibfield  [0]{\@secondoftwo}%
\providecommand \translation [1]{[#1]}%
\providecommand \BibitemOpen [0]{}%
\providecommand \bibitemStop [0]{}%
\providecommand \bibitemNoStop [0]{.\EOS\space}%
\providecommand \EOS [0]{\spacefactor3000\relax}%
\providecommand \BibitemShut  [1]{\csname bibitem#1\endcsname}%
\let\auto@bib@innerbib\@empty
\bibitem [{\citenamefont {Pfleiderer}(2009)}]{Pfleiderer2009}%
  \BibitemOpen
  \bibfield  {author} {\bibinfo {author} {\bibfnamefont {C.}~\bibnamefont
  {Pfleiderer}},\ }\bibfield  {title} {\bibinfo {title} {Superconducting phases
  of $f$-electron compounds},\ }\href
  {https://doi.org/10.1103/RevModPhys.81.1551} {\bibfield  {journal} {\bibinfo
  {journal} {Rev. Mod. Phys.}\ }\textbf {\bibinfo {volume} {81}},\ \bibinfo
  {pages} {1551} (\bibinfo {year} {2009})}\BibitemShut {NoStop}%
\bibitem [{\citenamefont {Hegger}\ \emph {et~al.}(2000)\citenamefont {Hegger},
  \citenamefont {Petrovic}, \citenamefont {Moshopoulou}, \citenamefont
  {Hundley}, \citenamefont {Sarrao}, \citenamefont {Fisk},\ and\ \citenamefont
  {Thompson}}]{Hegger2000}%
  \BibitemOpen
  \bibfield  {author} {\bibinfo {author} {\bibfnamefont {H.}~\bibnamefont
  {Hegger}}, \bibinfo {author} {\bibfnamefont {C.}~\bibnamefont {Petrovic}},
  \bibinfo {author} {\bibfnamefont {E.~G.}\ \bibnamefont {Moshopoulou}},
  \bibinfo {author} {\bibfnamefont {M.~F.}\ \bibnamefont {Hundley}}, \bibinfo
  {author} {\bibfnamefont {J.~L.}\ \bibnamefont {Sarrao}}, \bibinfo {author}
  {\bibfnamefont {Z.}~\bibnamefont {Fisk}},\ and\ \bibinfo {author}
  {\bibfnamefont {J.~D.}\ \bibnamefont {Thompson}},\ }\bibfield  {title}
  {\bibinfo {title} {Pressure-induced superconductivity in quasi-2{D}
  {CeRhIn}$_{5}$},\ }\href {https://doi.org/10.1103/PhysRevLett.84.4986}
  {\bibfield  {journal} {\bibinfo  {journal} {Phys. Rev. Lett.}\ }\textbf
  {\bibinfo {volume} {84}},\ \bibinfo {pages} {4986} (\bibinfo {year}
  {2000})}\BibitemShut {NoStop}%
\bibitem [{\citenamefont {Petrovic}\ \emph
  {et~al.}(2001{\natexlab{a}})\citenamefont {Petrovic}, \citenamefont
  {Movshovich}, \citenamefont {Jaime}, \citenamefont {Pagliuso}, \citenamefont
  {Hundley}, \citenamefont {Sarrao}, \citenamefont {Fisk},\ and\ \citenamefont
  {Thompson}}]{Petrovic2001CeIrIn5}%
  \BibitemOpen
  \bibfield  {author} {\bibinfo {author} {\bibfnamefont {C.}~\bibnamefont
  {Petrovic}}, \bibinfo {author} {\bibfnamefont {R.}~\bibnamefont
  {Movshovich}}, \bibinfo {author} {\bibfnamefont {M.}~\bibnamefont {Jaime}},
  \bibinfo {author} {\bibfnamefont {P.~G.}\ \bibnamefont {Pagliuso}}, \bibinfo
  {author} {\bibfnamefont {M.~F.}\ \bibnamefont {Hundley}}, \bibinfo {author}
  {\bibfnamefont {J.~L.}\ \bibnamefont {Sarrao}}, \bibinfo {author}
  {\bibfnamefont {Z.}~\bibnamefont {Fisk}},\ and\ \bibinfo {author}
  {\bibfnamefont {J.~D.}\ \bibnamefont {Thompson}},\ }\bibfield  {title}
  {\bibinfo {title} {A new heavy-fermion superconductor {CeIrIn}$_5$: A
  relative of the cuprates?},\ }\href
  {https://doi.org/10.1209/epl/i2001-00161-8} {\bibfield  {journal} {\bibinfo
  {journal} {Europhys. Lett.}\ }\textbf {\bibinfo {volume} {53}},\ \bibinfo
  {pages} {354} (\bibinfo {year} {2001}{\natexlab{a}})}\BibitemShut {NoStop}%
\bibitem [{\citenamefont {Petrovic}\ \emph
  {et~al.}(2001{\natexlab{b}})\citenamefont {Petrovic}, \citenamefont
  {Pagliuso}, \citenamefont {Hundley}, \citenamefont {Movshovich},
  \citenamefont {Sarrao}, \citenamefont {Thompson}, \citenamefont {Fisk},\ and\
  \citenamefont {Monthoux}}]{Petrovic2001CeCoIn5}%
  \BibitemOpen
  \bibfield  {author} {\bibinfo {author} {\bibfnamefont {C.}~\bibnamefont
  {Petrovic}}, \bibinfo {author} {\bibfnamefont {P.~G.}\ \bibnamefont
  {Pagliuso}}, \bibinfo {author} {\bibfnamefont {M.~F.}\ \bibnamefont
  {Hundley}}, \bibinfo {author} {\bibfnamefont {R.}~\bibnamefont {Movshovich}},
  \bibinfo {author} {\bibfnamefont {J.~L.}\ \bibnamefont {Sarrao}}, \bibinfo
  {author} {\bibfnamefont {J.~D.}\ \bibnamefont {Thompson}}, \bibinfo {author}
  {\bibfnamefont {Z.}~\bibnamefont {Fisk}},\ and\ \bibinfo {author}
  {\bibfnamefont {P.}~\bibnamefont {Monthoux}},\ }\bibfield  {title} {\bibinfo
  {title} {Heavy-fermion superconductivity in {CeCoIn}$_5$ at 2.3 {K}},\ }\href
  {http://stacks.iop.org/0953-8984/13/i=17/a=103} {\bibfield  {journal}
  {\bibinfo  {journal} {J. Phys.: Cond. Matt.}\ }\textbf {\bibinfo {volume}
  {13}},\ \bibinfo {pages} {L337} (\bibinfo {year}
  {2001}{\natexlab{b}})}\BibitemShut {NoStop}%
\bibitem [{\citenamefont {Thompson}\ and\ \citenamefont
  {Fisk}(2012)}]{Thompson2012}%
  \BibitemOpen
  \bibfield  {author} {\bibinfo {author} {\bibfnamefont {J.~D.}\ \bibnamefont
  {Thompson}}\ and\ \bibinfo {author} {\bibfnamefont {Z.}~\bibnamefont
  {Fisk}},\ }\bibfield  {title} {\bibinfo {title} {Progress in heavy-fermion
  superconductivity: Ce115 and related materials},\ }\href
  {https://doi.org/10.1143/JPSJ.81.011002} {\bibfield  {journal} {\bibinfo
  {journal} {J. Phys. Soc. Jpn.}\ }\textbf {\bibinfo {volume} {81}},\ \bibinfo
  {pages} {011002} (\bibinfo {year} {2012})}\BibitemShut {NoStop}%
\bibitem [{\citenamefont {Young}\ \emph {et~al.}(2007)\citenamefont {Young},
  \citenamefont {Urbano}, \citenamefont {Curro}, \citenamefont {Thompson},
  \citenamefont {Sarrao}, \citenamefont {Vorontsov},\ and\ \citenamefont
  {Graf}}]{Young2007}%
  \BibitemOpen
  \bibfield  {author} {\bibinfo {author} {\bibfnamefont {B.-L.}\ \bibnamefont
  {Young}}, \bibinfo {author} {\bibfnamefont {R.~R.}\ \bibnamefont {Urbano}},
  \bibinfo {author} {\bibfnamefont {N.~J.}\ \bibnamefont {Curro}}, \bibinfo
  {author} {\bibfnamefont {J.~D.}\ \bibnamefont {Thompson}}, \bibinfo {author}
  {\bibfnamefont {J.~L.}\ \bibnamefont {Sarrao}}, \bibinfo {author}
  {\bibfnamefont {A.~B.}\ \bibnamefont {Vorontsov}},\ and\ \bibinfo {author}
  {\bibfnamefont {M.~J.}\ \bibnamefont {Graf}},\ }\bibfield  {title} {\bibinfo
  {title} {Microscopic evidence for field-induced magnetism in
  {CeCoIn}$_{5}$},\ }\href {https://doi.org/10.1103/PhysRevLett.98.036402}
  {\bibfield  {journal} {\bibinfo  {journal} {Phys. Rev. Lett.}\ }\textbf
  {\bibinfo {volume} {98}},\ \bibinfo {pages} {036402} (\bibinfo {year}
  {2007})}\BibitemShut {NoStop}%
\bibitem [{\citenamefont {Gofryk}\ \emph {et~al.}(2012)\citenamefont {Gofryk},
  \citenamefont {Ronning}, \citenamefont {Zhu}, \citenamefont {Ou},
  \citenamefont {Tobash}, \citenamefont {Stoyko}, \citenamefont {Lu},
  \citenamefont {Mar}, \citenamefont {Park}, \citenamefont {Bauer},
  \citenamefont {Thompson},\ and\ \citenamefont {Fisk}}]{Gofryk2012}%
  \BibitemOpen
  \bibfield  {author} {\bibinfo {author} {\bibfnamefont {K.}~\bibnamefont
  {Gofryk}}, \bibinfo {author} {\bibfnamefont {F.}~\bibnamefont {Ronning}},
  \bibinfo {author} {\bibfnamefont {J.-X.}\ \bibnamefont {Zhu}}, \bibinfo
  {author} {\bibfnamefont {M.~N.}\ \bibnamefont {Ou}}, \bibinfo {author}
  {\bibfnamefont {P.~H.}\ \bibnamefont {Tobash}}, \bibinfo {author}
  {\bibfnamefont {S.~S.}\ \bibnamefont {Stoyko}}, \bibinfo {author}
  {\bibfnamefont {X.}~\bibnamefont {Lu}}, \bibinfo {author} {\bibfnamefont
  {A.}~\bibnamefont {Mar}}, \bibinfo {author} {\bibfnamefont {T.}~\bibnamefont
  {Park}}, \bibinfo {author} {\bibfnamefont {E.~D.}\ \bibnamefont {Bauer}},
  \bibinfo {author} {\bibfnamefont {J.~D.}\ \bibnamefont {Thompson}},\ and\
  \bibinfo {author} {\bibfnamefont {Z.}~\bibnamefont {Fisk}},\ }\bibfield
  {title} {\bibinfo {title} {Electronic tuning and uniform superconductivity in
  {CeCoIn}$_{5}$},\ }\href {https://doi.org/10.1103/PhysRevLett.109.186402}
  {\bibfield  {journal} {\bibinfo  {journal} {Phys. Rev. Lett.}\ }\textbf
  {\bibinfo {volume} {109}},\ \bibinfo {pages} {186402} (\bibinfo {year}
  {2012})}\BibitemShut {NoStop}%
\bibitem [{\citenamefont {Stock}\ \emph {et~al.}(2018)\citenamefont {Stock},
  \citenamefont {Rodriguez-Rivera}, \citenamefont {Schmalzl}, \citenamefont
  {Demmel}, \citenamefont {Singh}, \citenamefont {Ronning}, \citenamefont
  {Thompson},\ and\ \citenamefont {Bauer}}]{Stock2018}%
  \BibitemOpen
  \bibfield  {author} {\bibinfo {author} {\bibfnamefont {C.}~\bibnamefont
  {Stock}}, \bibinfo {author} {\bibfnamefont {J.~A.}\ \bibnamefont
  {Rodriguez-Rivera}}, \bibinfo {author} {\bibfnamefont {K.}~\bibnamefont
  {Schmalzl}}, \bibinfo {author} {\bibfnamefont {F.}~\bibnamefont {Demmel}},
  \bibinfo {author} {\bibfnamefont {D.~K.}\ \bibnamefont {Singh}}, \bibinfo
  {author} {\bibfnamefont {F.}~\bibnamefont {Ronning}}, \bibinfo {author}
  {\bibfnamefont {J.~D.}\ \bibnamefont {Thompson}},\ and\ \bibinfo {author}
  {\bibfnamefont {E.~D.}\ \bibnamefont {Bauer}},\ }\bibfield  {title} {\bibinfo
  {title} {From {I}sing resonant fluctuations to static uniaxial order in
  antiferromagnetic and weakly superconducting
  {CeCo}({In}$_{1-x}${Hg}$_{x}$)$_{5}$ (x=0.01)},\ }\href
  {https://doi.org/10.1103/PhysRevLett.121.037003} {\bibfield  {journal}
  {\bibinfo  {journal} {Phys. Rev. Lett.}\ }\textbf {\bibinfo {volume} {121}},\
  \bibinfo {pages} {037003} (\bibinfo {year} {2018})}\BibitemShut {NoStop}%
\bibitem [{\citenamefont {Mathur}\ \emph {et~al.}(1998)\citenamefont {Mathur},
  \citenamefont {Grosche}, \citenamefont {Julian}, \citenamefont {Walker},
  \citenamefont {Freye}, \citenamefont {Haselwimmer},\ and\ \citenamefont
  {Lonzarich}}]{Mathur1998}%
  \BibitemOpen
  \bibfield  {author} {\bibinfo {author} {\bibfnamefont {N.~D.}\ \bibnamefont
  {Mathur}}, \bibinfo {author} {\bibfnamefont {F.~M.}\ \bibnamefont {Grosche}},
  \bibinfo {author} {\bibfnamefont {S.~R.}\ \bibnamefont {Julian}}, \bibinfo
  {author} {\bibfnamefont {I.~R.}\ \bibnamefont {Walker}}, \bibinfo {author}
  {\bibfnamefont {D.~M.}\ \bibnamefont {Freye}}, \bibinfo {author}
  {\bibfnamefont {R.~K.~W.}\ \bibnamefont {Haselwimmer}},\ and\ \bibinfo
  {author} {\bibfnamefont {G.~G.}\ \bibnamefont {Lonzarich}},\ }\bibfield
  {title} {\bibinfo {title} {Magnetically mediated superconductivity in heavy
  fermion compounds},\ }\href {https://doi.org/10.1038/27838} {\bibfield
  {journal} {\bibinfo  {journal} {Nature}\ }\textbf {\bibinfo {volume} {394}},\
  \bibinfo {pages} {39} (\bibinfo {year} {1998})}\BibitemShut {NoStop}%
\bibitem [{\citenamefont {Kohori}\ \emph {et~al.}(2001)\citenamefont {Kohori},
  \citenamefont {Yamato}, \citenamefont {Iwamoto}, \citenamefont {Kohara},
  \citenamefont {Bauer}, \citenamefont {Maple},\ and\ \citenamefont
  {Sarrao}}]{Kohori2001}%
  \BibitemOpen
  \bibfield  {author} {\bibinfo {author} {\bibfnamefont {Y.}~\bibnamefont
  {Kohori}}, \bibinfo {author} {\bibfnamefont {Y.}~\bibnamefont {Yamato}},
  \bibinfo {author} {\bibfnamefont {Y.}~\bibnamefont {Iwamoto}}, \bibinfo
  {author} {\bibfnamefont {T.}~\bibnamefont {Kohara}}, \bibinfo {author}
  {\bibfnamefont {E.~D.}\ \bibnamefont {Bauer}}, \bibinfo {author}
  {\bibfnamefont {M.~B.}\ \bibnamefont {Maple}},\ and\ \bibinfo {author}
  {\bibfnamefont {J.~L.}\ \bibnamefont {Sarrao}},\ }\bibfield  {title}
  {\bibinfo {title} {{NMR} and {NQR} studies of the heavy fermion
  superconductors {CeTIn}$_{5}$ ({T}={Co} and {Ir})},\ }\href
  {https://doi.org/10.1103/PhysRevB.64.134526} {\bibfield  {journal} {\bibinfo
  {journal} {Phys. Rev. B}\ }\textbf {\bibinfo {volume} {64}},\ \bibinfo
  {pages} {134526} (\bibinfo {year} {2001})}\BibitemShut {NoStop}%
\bibitem [{\citenamefont {Movshovich}\ \emph {et~al.}(2001)\citenamefont
  {Movshovich}, \citenamefont {Jaime}, \citenamefont {Thompson}, \citenamefont
  {Petrovic}, \citenamefont {Fisk}, \citenamefont {Pagliuso},\ and\
  \citenamefont {Sarrao}}]{Movshovich2001}%
  \BibitemOpen
  \bibfield  {author} {\bibinfo {author} {\bibfnamefont {R.}~\bibnamefont
  {Movshovich}}, \bibinfo {author} {\bibfnamefont {M.}~\bibnamefont {Jaime}},
  \bibinfo {author} {\bibfnamefont {J.~D.}\ \bibnamefont {Thompson}}, \bibinfo
  {author} {\bibfnamefont {C.}~\bibnamefont {Petrovic}}, \bibinfo {author}
  {\bibfnamefont {Z.}~\bibnamefont {Fisk}}, \bibinfo {author} {\bibfnamefont
  {P.~G.}\ \bibnamefont {Pagliuso}},\ and\ \bibinfo {author} {\bibfnamefont
  {J.~L.}\ \bibnamefont {Sarrao}},\ }\bibfield  {title} {\bibinfo {title}
  {Unconventional superconductivity in {CeIrIn}$_{5}$ and {CeCoIn}$_{5}$:
  Specific heat and thermal conductivity studies},\ }\href
  {https://doi.org/10.1103/PhysRevLett.86.5152} {\bibfield  {journal} {\bibinfo
   {journal} {Phys. Rev. Lett.}\ }\textbf {\bibinfo {volume} {86}},\ \bibinfo
  {pages} {5152} (\bibinfo {year} {2001})}\BibitemShut {NoStop}%
\bibitem [{\citenamefont {{\"O}zcan}\ \emph {et~al.}(2003)\citenamefont
  {{\"O}zcan}, \citenamefont {Broun}, \citenamefont {Morgan}, \citenamefont
  {Haselwimmer}, \citenamefont {Sarrao}, \citenamefont {Kamal}, \citenamefont
  {Bidinosti}, \citenamefont {Turner}, \citenamefont {Raudsepp},\ and\
  \citenamefont {Waldram}}]{Ozcan2003}%
  \BibitemOpen
  \bibfield  {author} {\bibinfo {author} {\bibfnamefont {S.}~\bibnamefont
  {{\"O}zcan}}, \bibinfo {author} {\bibfnamefont {D.~M.}\ \bibnamefont
  {Broun}}, \bibinfo {author} {\bibfnamefont {B.}~\bibnamefont {Morgan}},
  \bibinfo {author} {\bibfnamefont {R.~K.~W.}\ \bibnamefont {Haselwimmer}},
  \bibinfo {author} {\bibfnamefont {J.~L.}\ \bibnamefont {Sarrao}}, \bibinfo
  {author} {\bibfnamefont {S.}~\bibnamefont {Kamal}}, \bibinfo {author}
  {\bibfnamefont {C.~P.}\ \bibnamefont {Bidinosti}}, \bibinfo {author}
  {\bibfnamefont {P.~J.}\ \bibnamefont {Turner}}, \bibinfo {author}
  {\bibfnamefont {M.}~\bibnamefont {Raudsepp}},\ and\ \bibinfo {author}
  {\bibfnamefont {J.~R.}\ \bibnamefont {Waldram}},\ }\bibfield  {title}
  {\bibinfo {title} {London penetration depth measurements of the heavy-fermion
  superconductor cecoin5 near a magnetic quantum critical point},\ }\href
  {https://doi.org/10.1209/epl/i2003-00411-9} {\bibfield  {journal} {\bibinfo
  {journal} {Europhysics Letters}\ }\textbf {\bibinfo {volume} {62}},\ \bibinfo
  {pages} {412} (\bibinfo {year} {2003})}\BibitemShut {NoStop}%
\bibitem [{\citenamefont {Rourke}\ \emph {et~al.}(2005)\citenamefont {Rourke},
  \citenamefont {Tanatar}, \citenamefont {Turel}, \citenamefont {Berdeklis},
  \citenamefont {Petrovic},\ and\ \citenamefont {Wei}}]{Rourke2005}%
  \BibitemOpen
  \bibfield  {author} {\bibinfo {author} {\bibfnamefont {P.~M.~C.}\
  \bibnamefont {Rourke}}, \bibinfo {author} {\bibfnamefont {M.~A.}\
  \bibnamefont {Tanatar}}, \bibinfo {author} {\bibfnamefont {C.~S.}\
  \bibnamefont {Turel}}, \bibinfo {author} {\bibfnamefont {J.}~\bibnamefont
  {Berdeklis}}, \bibinfo {author} {\bibfnamefont {C.}~\bibnamefont
  {Petrovic}},\ and\ \bibinfo {author} {\bibfnamefont {J.~Y.~T.}\ \bibnamefont
  {Wei}},\ }\bibfield  {title} {\bibinfo {title} {Spectroscopic evidence for
  multiple order parameter components in the heavy fermion superconductor
  ${\mathrm{c}\mathrm{e}\mathrm{c}\mathrm{o}\mathrm{i}\mathrm{n}}_{5}$},\
  }\href {https://doi.org/10.1103/PhysRevLett.94.107005} {\bibfield  {journal}
  {\bibinfo  {journal} {Phys. Rev. Lett.}\ }\textbf {\bibinfo {volume} {94}},\
  \bibinfo {pages} {107005} (\bibinfo {year} {2005})}\BibitemShut {NoStop}%
\bibitem [{\citenamefont {Kim}\ \emph {et~al.}(2015)\citenamefont {Kim},
  \citenamefont {Tanatar}, \citenamefont {Flint}, \citenamefont {Petrovic},
  \citenamefont {Hu}, \citenamefont {White}, \citenamefont {Lum}, \citenamefont
  {Maple},\ and\ \citenamefont {Prozorov}}]{Kim2015}%
  \BibitemOpen
  \bibfield  {author} {\bibinfo {author} {\bibfnamefont {H.}~\bibnamefont
  {Kim}}, \bibinfo {author} {\bibfnamefont {M.~A.}\ \bibnamefont {Tanatar}},
  \bibinfo {author} {\bibfnamefont {R.}~\bibnamefont {Flint}}, \bibinfo
  {author} {\bibfnamefont {C.}~\bibnamefont {Petrovic}}, \bibinfo {author}
  {\bibfnamefont {R.}~\bibnamefont {Hu}}, \bibinfo {author} {\bibfnamefont
  {B.~D.}\ \bibnamefont {White}}, \bibinfo {author} {\bibfnamefont {I.~K.}\
  \bibnamefont {Lum}}, \bibinfo {author} {\bibfnamefont {M.~B.}\ \bibnamefont
  {Maple}},\ and\ \bibinfo {author} {\bibfnamefont {R.}~\bibnamefont
  {Prozorov}},\ }\bibfield  {title} {\bibinfo {title} {Nodal to nodeless
  superconducting energy-gap structure change concomitant with fermi-surface
  reconstruction in the heavy-fermion compound {CeCoIn}$_{5}$},\ }\href
  {https://doi.org/10.1103/PhysRevLett.114.027003} {\bibfield  {journal}
  {\bibinfo  {journal} {Phys. Rev. Lett.}\ }\textbf {\bibinfo {volume} {114}},\
  \bibinfo {pages} {027003} (\bibinfo {year} {2015})}\BibitemShut {NoStop}%
\bibitem [{\citenamefont {Allan}\ \emph {et~al.}(2013)\citenamefont {Allan},
  \citenamefont {Massee}, \citenamefont {Morr}, \citenamefont {Van~Dyke},
  \citenamefont {Rost}, \citenamefont {Mackenzie}, \citenamefont {Petrovic},\
  and\ \citenamefont {Davis}}]{Allan2013}%
  \BibitemOpen
  \bibfield  {author} {\bibinfo {author} {\bibfnamefont {M.~P.}\ \bibnamefont
  {Allan}}, \bibinfo {author} {\bibfnamefont {F.}~\bibnamefont {Massee}},
  \bibinfo {author} {\bibfnamefont {D.~K.}\ \bibnamefont {Morr}}, \bibinfo
  {author} {\bibfnamefont {J.}~\bibnamefont {Van~Dyke}}, \bibinfo {author}
  {\bibfnamefont {A.~W.}\ \bibnamefont {Rost}}, \bibinfo {author}
  {\bibfnamefont {A.~P.}\ \bibnamefont {Mackenzie}}, \bibinfo {author}
  {\bibfnamefont {C.}~\bibnamefont {Petrovic}},\ and\ \bibinfo {author}
  {\bibfnamefont {J.~C.}\ \bibnamefont {Davis}},\ }\bibfield  {title} {\bibinfo
  {title} {Imaging cooper pairing of heavy fermions in {CeCoIn}$_5$},\ }\href
  {https://doi.org/10.1038/nphys2671} {\bibfield  {journal} {\bibinfo
  {journal} {Nature Phys.}\ }\textbf {\bibinfo {volume} {9}},\ \bibinfo {pages}
  {468} (\bibinfo {year} {2013})}\BibitemShut {NoStop}%
\bibitem [{\citenamefont {Zhou}\ \emph {et~al.}(2013)\citenamefont {Zhou},
  \citenamefont {Misra}, \citenamefont {da~Silva~Neto}, \citenamefont
  {Aynajian}, \citenamefont {Baumbach}, \citenamefont {Thompson}, \citenamefont
  {Bauer},\ and\ \citenamefont {Yazdani}}]{Zhou2013}%
  \BibitemOpen
  \bibfield  {author} {\bibinfo {author} {\bibfnamefont {B.~B.}\ \bibnamefont
  {Zhou}}, \bibinfo {author} {\bibfnamefont {S.}~\bibnamefont {Misra}},
  \bibinfo {author} {\bibfnamefont {E.~H.}\ \bibnamefont {da~Silva~Neto}},
  \bibinfo {author} {\bibfnamefont {P.}~\bibnamefont {Aynajian}}, \bibinfo
  {author} {\bibfnamefont {R.~E.}\ \bibnamefont {Baumbach}}, \bibinfo {author}
  {\bibfnamefont {J.~D.}\ \bibnamefont {Thompson}}, \bibinfo {author}
  {\bibfnamefont {E.~D.}\ \bibnamefont {Bauer}},\ and\ \bibinfo {author}
  {\bibfnamefont {A.}~\bibnamefont {Yazdani}},\ }\bibfield  {title} {\bibinfo
  {title} {Visualizing nodal heavy fermion superconductivity in {CeCoIn}$_5$},\
  }\href {https://doi.org/10.1038/nphys2672} {\bibfield  {journal} {\bibinfo
  {journal} {Nature Phys.}\ }\textbf {\bibinfo {volume} {9}},\ \bibinfo {pages}
  {474} (\bibinfo {year} {2013})}\BibitemShut {NoStop}%
\bibitem [{\citenamefont {Izawa}\ \emph {et~al.}(2001)\citenamefont {Izawa},
  \citenamefont {Yamaguchi}, \citenamefont {Matsuda}, \citenamefont {Shishido},
  \citenamefont {Settai},\ and\ \citenamefont {Onuki}}]{Izawa2001}%
  \BibitemOpen
  \bibfield  {author} {\bibinfo {author} {\bibfnamefont {K.}~\bibnamefont
  {Izawa}}, \bibinfo {author} {\bibfnamefont {H.}~\bibnamefont {Yamaguchi}},
  \bibinfo {author} {\bibfnamefont {Y.}~\bibnamefont {Matsuda}}, \bibinfo
  {author} {\bibfnamefont {H.}~\bibnamefont {Shishido}}, \bibinfo {author}
  {\bibfnamefont {R.}~\bibnamefont {Settai}},\ and\ \bibinfo {author}
  {\bibfnamefont {Y.}~\bibnamefont {Onuki}},\ }\bibfield  {title} {\bibinfo
  {title} {Angular position of nodes in the superconducting gap of quasi-2{D}
  heavy-fermion superconductor {CeCoIn}$_{5}$},\ }\href
  {https://doi.org/10.1103/PhysRevLett.87.057002} {\bibfield  {journal}
  {\bibinfo  {journal} {Phys. Rev. Lett.}\ }\textbf {\bibinfo {volume} {87}},\
  \bibinfo {pages} {057002} (\bibinfo {year} {2001})}\BibitemShut {NoStop}%
\bibitem [{\citenamefont {Vorontsov}\ and\ \citenamefont
  {Vekhter}(2006)}]{Vorontsov2006}%
  \BibitemOpen
  \bibfield  {author} {\bibinfo {author} {\bibfnamefont {A.}~\bibnamefont
  {Vorontsov}}\ and\ \bibinfo {author} {\bibfnamefont {I.}~\bibnamefont
  {Vekhter}},\ }\bibfield  {title} {\bibinfo {title} {Nodal structure of
  quasi-two-dimensional superconductors probed by a magnetic field},\ }\href
  {https://doi.org/10.1103/PhysRevLett.96.237001} {\bibfield  {journal}
  {\bibinfo  {journal} {Phys. Rev. Lett.}\ }\textbf {\bibinfo {volume} {96}},\
  \bibinfo {pages} {237001} (\bibinfo {year} {2006})}\BibitemShut {NoStop}%
\bibitem [{\citenamefont {Aoki}\ \emph {et~al.}(2004)\citenamefont {Aoki},
  \citenamefont {Sakakibara}, \citenamefont {Shishido}, \citenamefont {Settai},
  \citenamefont {Onuki}, \citenamefont {Miranovi\'c},\ and\ \citenamefont
  {Machida}}]{Aoki2004}%
  \BibitemOpen
  \bibfield  {author} {\bibinfo {author} {\bibfnamefont {H.}~\bibnamefont
  {Aoki}}, \bibinfo {author} {\bibfnamefont {T.}~\bibnamefont {Sakakibara}},
  \bibinfo {author} {\bibfnamefont {H.}~\bibnamefont {Shishido}}, \bibinfo
  {author} {\bibfnamefont {R.}~\bibnamefont {Settai}}, \bibinfo {author}
  {\bibfnamefont {Y.}~\bibnamefont {Onuki}}, \bibinfo {author} {\bibfnamefont
  {P.}~\bibnamefont {Miranovi\'c}},\ and\ \bibinfo {author} {\bibfnamefont
  {K.}~\bibnamefont {Machida}},\ }\bibfield  {title} {\bibinfo {title}
  {Field-angle dependence of the zero-energy density of states in the
  unconventional heavy-fermion superconductor {CeCoIn}$_5$},\ }\href
  {https://doi.org/10.1088/0953-8984/16/3/l02} {\bibfield  {journal} {\bibinfo
  {journal} {J. Phys.: Cond. Matt.}\ }\textbf {\bibinfo {volume} {16}},\
  \bibinfo {pages} {L13} (\bibinfo {year} {2004})}\BibitemShut {NoStop}%
\bibitem [{\citenamefont {Stock}\ \emph {et~al.}(2008)\citenamefont {Stock},
  \citenamefont {Broholm}, \citenamefont {Hudis}, \citenamefont {Kang},\ and\
  \citenamefont {Petrovic}}]{Stock2008}%
  \BibitemOpen
  \bibfield  {author} {\bibinfo {author} {\bibfnamefont {C.}~\bibnamefont
  {Stock}}, \bibinfo {author} {\bibfnamefont {C.}~\bibnamefont {Broholm}},
  \bibinfo {author} {\bibfnamefont {J.}~\bibnamefont {Hudis}}, \bibinfo
  {author} {\bibfnamefont {H.~J.}\ \bibnamefont {Kang}},\ and\ \bibinfo
  {author} {\bibfnamefont {C.}~\bibnamefont {Petrovic}},\ }\bibfield  {title}
  {\bibinfo {title} {Spin resonance in the $d$-wave superconductor
  {CeCoIn}$_{5}$},\ }\href {https://doi.org/10.1103/PhysRevLett.100.087001}
  {\bibfield  {journal} {\bibinfo  {journal} {Phys. Rev. Lett.}\ }\textbf
  {\bibinfo {volume} {100}},\ \bibinfo {pages} {087001} (\bibinfo {year}
  {2008})}\BibitemShut {NoStop}%
\bibitem [{\citenamefont {Gu}\ \emph {et~al.}(2019)\citenamefont {Gu},
  \citenamefont {Wan}, \citenamefont {Tang}, \citenamefont {Du}, \citenamefont
  {Yang}, \citenamefont {Wang}, \citenamefont {Zhong}, \citenamefont {Wen},
  \citenamefont {Gu},\ and\ \citenamefont {Wen}}]{Gu2019}%
  \BibitemOpen
  \bibfield  {author} {\bibinfo {author} {\bibfnamefont {Q.}~\bibnamefont
  {Gu}}, \bibinfo {author} {\bibfnamefont {S.}~\bibnamefont {Wan}}, \bibinfo
  {author} {\bibfnamefont {Q.}~\bibnamefont {Tang}}, \bibinfo {author}
  {\bibfnamefont {Z.}~\bibnamefont {Du}}, \bibinfo {author} {\bibfnamefont
  {H.}~\bibnamefont {Yang}}, \bibinfo {author} {\bibfnamefont {Q.-H.}\
  \bibnamefont {Wang}}, \bibinfo {author} {\bibfnamefont {R.}~\bibnamefont
  {Zhong}}, \bibinfo {author} {\bibfnamefont {J.}~\bibnamefont {Wen}}, \bibinfo
  {author} {\bibfnamefont {G.~D.}\ \bibnamefont {Gu}},\ and\ \bibinfo {author}
  {\bibfnamefont {H.-H.}\ \bibnamefont {Wen}},\ }\bibfield  {title} {\bibinfo
  {title} {Directly visualizing the sign change of $d$-wave superconducting gap
  in {Bi}$_2${Sr}$_2${CaCu}$_2${O}$_{8+\delta}$ by phase-referenced
  quasiparticle interference},\ }\href
  {https://doi.org/10.1038/s41467-019-09340-5} {\bibfield  {journal} {\bibinfo
  {journal} {Nature Comm.}\ }\textbf {\bibinfo {volume} {10}},\ \bibinfo
  {pages} {1603} (\bibinfo {year} {2019})}\BibitemShut {NoStop}%
\bibitem [{\citenamefont {Chubukov}\ \emph {et~al.}(2008)\citenamefont
  {Chubukov}, \citenamefont {Pines},\ and\ \citenamefont
  {Schmalian}}]{Chubukov2008}%
  \BibitemOpen
  \bibfield  {author} {\bibinfo {author} {\bibfnamefont {A.~V.}\ \bibnamefont
  {Chubukov}}, \bibinfo {author} {\bibfnamefont {D.}~\bibnamefont {Pines}},\
  and\ \bibinfo {author} {\bibfnamefont {J.}~\bibnamefont {Schmalian}},\
  }\bibinfo {title} {A spin fluctuation model for d-wave superconductivity},\
  in\ \href {https://doi.org/10.1007/978-3-540-73253-2{\_}22} {\emph {\bibinfo
  {booktitle} {Superconductivity: Conventional and Unconventional
  Superconductors}}},\ \bibinfo {editor} {edited by\ \bibinfo {editor}
  {\bibfnamefont {K.~H.}\ \bibnamefont {Bennemann}}\ and\ \bibinfo {editor}
  {\bibfnamefont {J.~B.}\ \bibnamefont {Ketterson}}}\ (\bibinfo  {publisher}
  {Springer Berlin Heidelberg},\ \bibinfo {address} {Berlin, Heidelberg},\
  \bibinfo {year} {2008})\ pp.\ \bibinfo {pages} {1349--1413}\BibitemShut
  {NoStop}%
\bibitem [{\citenamefont {Kogan}\ \emph {et~al.}(2009)\citenamefont {Kogan},
  \citenamefont {Prozorov},\ and\ \citenamefont {Petrovic}}]{Kogan2009}%
  \BibitemOpen
  \bibfield  {author} {\bibinfo {author} {\bibfnamefont {V.~G.}\ \bibnamefont
  {Kogan}}, \bibinfo {author} {\bibfnamefont {R.}~\bibnamefont {Prozorov}},\
  and\ \bibinfo {author} {\bibfnamefont {C.}~\bibnamefont {Petrovic}},\
  }\bibfield  {title} {\bibinfo {title} {Superfluid density in gapless
  superconductor {CeCoIn}$_5$},\ }\href
  {http://stacks.iop.org/0953-8984/21/i=10/a=102204} {\bibfield  {journal}
  {\bibinfo  {journal} {J. Phys.: Cond. Matt.}\ }\textbf {\bibinfo {volume}
  {21}},\ \bibinfo {pages} {102204} (\bibinfo {year} {2009})}\BibitemShut
  {NoStop}%
\bibitem [{\citenamefont {Howald}\ \emph {et~al.}(2013)\citenamefont {Howald},
  \citenamefont {Maisuradze}, \citenamefont {de~R\'eotier}, \citenamefont
  {Yaouanc}, \citenamefont {Baines}, \citenamefont {Lapertot}, \citenamefont
  {Mony}, \citenamefont {Brison},\ and\ \citenamefont {Keller}}]{Howald2013}%
  \BibitemOpen
  \bibfield  {author} {\bibinfo {author} {\bibfnamefont {L.}~\bibnamefont
  {Howald}}, \bibinfo {author} {\bibfnamefont {A.}~\bibnamefont {Maisuradze}},
  \bibinfo {author} {\bibfnamefont {P.~D.}\ \bibnamefont {de~R\'eotier}},
  \bibinfo {author} {\bibfnamefont {A.}~\bibnamefont {Yaouanc}}, \bibinfo
  {author} {\bibfnamefont {C.}~\bibnamefont {Baines}}, \bibinfo {author}
  {\bibfnamefont {G.}~\bibnamefont {Lapertot}}, \bibinfo {author}
  {\bibfnamefont {K.}~\bibnamefont {Mony}}, \bibinfo {author} {\bibfnamefont
  {J.-P.}\ \bibnamefont {Brison}},\ and\ \bibinfo {author} {\bibfnamefont
  {H.}~\bibnamefont {Keller}},\ }\bibfield  {title} {\bibinfo {title} {Strong
  pressure dependence of the magnetic penetration depth in single crystals of
  the heavy-fermion superconductor {CeCoIn}$_{5}$ studied by muon spin
  rotation},\ }\href {https://doi.org/10.1103/PhysRevLett.110.017005}
  {\bibfield  {journal} {\bibinfo  {journal} {Phys. Rev. Lett.}\ }\textbf
  {\bibinfo {volume} {110}},\ \bibinfo {pages} {017005} (\bibinfo {year}
  {2013})}\BibitemShut {NoStop}%
\bibitem [{\citenamefont {Amin}\ \emph {et~al.}(1998)\citenamefont {Amin},
  \citenamefont {Affleck},\ and\ \citenamefont {Franz}}]{Amin1998}%
  \BibitemOpen
  \bibfield  {author} {\bibinfo {author} {\bibfnamefont {M.~H.~S.}\
  \bibnamefont {Amin}}, \bibinfo {author} {\bibfnamefont {I.}~\bibnamefont
  {Affleck}},\ and\ \bibinfo {author} {\bibfnamefont {M.}~\bibnamefont
  {Franz}},\ }\bibfield  {title} {\bibinfo {title} {Low-temperature behavior of
  the vortex lattice in unconventional superconductors},\ }\href
  {https://doi.org/10.1103/PhysRevB.58.5848} {\bibfield  {journal} {\bibinfo
  {journal} {Phys. Rev. B}\ }\textbf {\bibinfo {volume} {58}},\ \bibinfo
  {pages} {5848} (\bibinfo {year} {1998})}\BibitemShut {NoStop}%
\bibitem [{\citenamefont {Ichioka}\ \emph {et~al.}(1999)\citenamefont
  {Ichioka}, \citenamefont {Hasegawa},\ and\ \citenamefont
  {Machida}}]{Ichioka1999}%
  \BibitemOpen
  \bibfield  {author} {\bibinfo {author} {\bibfnamefont {M.}~\bibnamefont
  {Ichioka}}, \bibinfo {author} {\bibfnamefont {A.}~\bibnamefont {Hasegawa}},\
  and\ \bibinfo {author} {\bibfnamefont {K.}~\bibnamefont {Machida}},\
  }\bibfield  {title} {\bibinfo {title} {Field dependence of the vortex
  structure in d-wave and s-wave superconductors},\ }\href
  {https://doi.org/10.1103/PhysRevB.59.8902} {\bibfield  {journal} {\bibinfo
  {journal} {Phys. Rev. B}\ }\textbf {\bibinfo {volume} {59}},\ \bibinfo
  {pages} {8902} (\bibinfo {year} {1999})}\BibitemShut {NoStop}%
\bibitem [{\citenamefont {Abrikosov}(2017)}]{Abrikosov2017}%
  \BibitemOpen
  \bibfield  {author} {\bibinfo {author} {\bibfnamefont {A.}~\bibnamefont
  {Abrikosov}},\ }\href {https://books.google.com/books?id=tTo2DwAAQBAJ} {\emph
  {\bibinfo {title} {Fundamentals of the Theory of Metals}}}\ (\bibinfo
  {publisher} {Dover Publications},\ \bibinfo {year} {2017})\BibitemShut
  {NoStop}%
\bibitem [{\citenamefont {Cribier}\ \emph {et~al.}(1964)\citenamefont
  {Cribier}, \citenamefont {Jacrot}, \citenamefont {{Madhav Rao}},\ and\
  \citenamefont {Farnoux}}]{Cribier1964}%
  \BibitemOpen
  \bibfield  {author} {\bibinfo {author} {\bibfnamefont {D.}~\bibnamefont
  {Cribier}}, \bibinfo {author} {\bibfnamefont {B.}~\bibnamefont {Jacrot}},
  \bibinfo {author} {\bibfnamefont {L.}~\bibnamefont {{Madhav Rao}}},\ and\
  \bibinfo {author} {\bibfnamefont {B.}~\bibnamefont {Farnoux}},\ }\bibfield
  {title} {\bibinfo {title} {Mise en evidence par diffraction de neutrons d'une
  structure periodique du champ magnetique dans le niobium supraconducteur},\
  }\href {https://doi.org/https://doi.org/10.1016/0031-9163(64)90096-4}
  {\bibfield  {journal} {\bibinfo  {journal} {Phys. Lett.}\ }\textbf {\bibinfo
  {volume} {9}},\ \bibinfo {pages} {106} (\bibinfo {year} {1964})}\BibitemShut
  {NoStop}%
\bibitem [{\citenamefont {Essmann}\ and\ \citenamefont
  {Tr{\"a}uble}(1967)}]{Essmann1967}%
  \BibitemOpen
  \bibfield  {author} {\bibinfo {author} {\bibfnamefont {U.}~\bibnamefont
  {Essmann}}\ and\ \bibinfo {author} {\bibfnamefont {H.}~\bibnamefont
  {Tr{\"a}uble}},\ }\bibfield  {title} {\bibinfo {title} {The direct
  observation of individual flux lines in type {II} superconductors},\ }\href
  {https://doi.org/https://doi.org/10.1016/0375-9601(67)90819-5} {\bibfield
  {journal} {\bibinfo  {journal} {Phys. Lett. A}\ }\textbf {\bibinfo {volume}
  {24}},\ \bibinfo {pages} {526} (\bibinfo {year} {1967})}\BibitemShut
  {NoStop}%
\bibitem [{\citenamefont {Hoffmann}\ \emph {et~al.}(2022)\citenamefont
  {Hoffmann}, \citenamefont {Schlegel}, \citenamefont {Salazar}, \citenamefont
  {Sykora}, \citenamefont {Nag}, \citenamefont {Khanenko}, \citenamefont
  {Beck}, \citenamefont {Aswartham}, \citenamefont {Wurmehl}, \citenamefont
  {B\"uchner}, \citenamefont {Fasano},\ and\ \citenamefont
  {Hess}}]{Hoffmann2022}%
  \BibitemOpen
  \bibfield  {author} {\bibinfo {author} {\bibfnamefont {S.}~\bibnamefont
  {Hoffmann}}, \bibinfo {author} {\bibfnamefont {R.}~\bibnamefont {Schlegel}},
  \bibinfo {author} {\bibfnamefont {C.}~\bibnamefont {Salazar}}, \bibinfo
  {author} {\bibfnamefont {S.}~\bibnamefont {Sykora}}, \bibinfo {author}
  {\bibfnamefont {P.~K.}\ \bibnamefont {Nag}}, \bibinfo {author} {\bibfnamefont
  {P.}~\bibnamefont {Khanenko}}, \bibinfo {author} {\bibfnamefont
  {R.}~\bibnamefont {Beck}}, \bibinfo {author} {\bibfnamefont {S.}~\bibnamefont
  {Aswartham}}, \bibinfo {author} {\bibfnamefont {S.}~\bibnamefont {Wurmehl}},
  \bibinfo {author} {\bibfnamefont {B.}~\bibnamefont {B\"uchner}}, \bibinfo
  {author} {\bibfnamefont {Y.}~\bibnamefont {Fasano}},\ and\ \bibinfo {author}
  {\bibfnamefont {C.}~\bibnamefont {Hess}},\ }\bibfield  {title} {\bibinfo
  {title} {Absence of hexagonal-to-square lattice transition in {LiFeAs} vortex
  matter},\ }\href {https://doi.org/10.1103/PhysRevB.106.134507} {\bibfield
  {journal} {\bibinfo  {journal} {Phys. Rev. B}\ }\textbf {\bibinfo {volume}
  {106}},\ \bibinfo {pages} {134507} (\bibinfo {year} {2022})}\BibitemShut
  {NoStop}%
\bibitem [{\citenamefont {Kim}\ \emph {et~al.}(2011)\citenamefont {Kim},
  \citenamefont {Tanatar}, \citenamefont {Song}, \citenamefont {Kwon},\ and\
  \citenamefont {Prozorov}}]{Kim2011}%
  \BibitemOpen
  \bibfield  {author} {\bibinfo {author} {\bibfnamefont {H.}~\bibnamefont
  {Kim}}, \bibinfo {author} {\bibfnamefont {M.~A.}\ \bibnamefont {Tanatar}},
  \bibinfo {author} {\bibfnamefont {Y.~J.}\ \bibnamefont {Song}}, \bibinfo
  {author} {\bibfnamefont {Y.~S.}\ \bibnamefont {Kwon}},\ and\ \bibinfo
  {author} {\bibfnamefont {R.}~\bibnamefont {Prozorov}},\ }\bibfield  {title}
  {\bibinfo {title} {Nodeless two-gap superconducting state in single crystals
  of the stoichiometric iron pnictide {LiFeAs}},\ }\href
  {https://doi.org/10.1103/PhysRevB.83.100502} {\bibfield  {journal} {\bibinfo
  {journal} {Phys. Rev. B}\ }\textbf {\bibinfo {volume} {83}},\ \bibinfo
  {pages} {100502} (\bibinfo {year} {2011})}\BibitemShut {NoStop}%
\bibitem [{\citenamefont {Das}\ \emph {et~al.}(2012{\natexlab{a}})\citenamefont
  {Das}, \citenamefont {Rastovski}, \citenamefont {O'Brien}, \citenamefont
  {Schlesinger}, \citenamefont {Dewhurst}, \citenamefont {DeBeer-Schmitt},
  \citenamefont {Zhigadlo}, \citenamefont {Karpinski},\ and\ \citenamefont
  {Eskildsen}}]{Das2012MgB2}%
  \BibitemOpen
  \bibfield  {author} {\bibinfo {author} {\bibfnamefont {P.}~\bibnamefont
  {Das}}, \bibinfo {author} {\bibfnamefont {C.}~\bibnamefont {Rastovski}},
  \bibinfo {author} {\bibfnamefont {T.~R.}\ \bibnamefont {O'Brien}}, \bibinfo
  {author} {\bibfnamefont {K.~J.}\ \bibnamefont {Schlesinger}}, \bibinfo
  {author} {\bibfnamefont {C.~D.}\ \bibnamefont {Dewhurst}}, \bibinfo {author}
  {\bibfnamefont {L.}~\bibnamefont {DeBeer-Schmitt}}, \bibinfo {author}
  {\bibfnamefont {N.~D.}\ \bibnamefont {Zhigadlo}}, \bibinfo {author}
  {\bibfnamefont {J.}~\bibnamefont {Karpinski}},\ and\ \bibinfo {author}
  {\bibfnamefont {M.~R.}\ \bibnamefont {Eskildsen}},\ }\bibfield  {title}
  {\bibinfo {title} {Observation of well-ordered metastable vortex lattice
  phases in superconducting {MgB}$_{2}$ using small-angle neutron scattering},\
  }\href {https://doi.org/10.1103/PhysRevLett.108.167001} {\bibfield  {journal}
  {\bibinfo  {journal} {Phys. Rev. Lett.}\ }\textbf {\bibinfo {volume} {108}},\
  \bibinfo {pages} {167001} (\bibinfo {year} {2012}{\natexlab{a}})}\BibitemShut
  {NoStop}%
\bibitem [{\citenamefont {Hirano}\ \emph {et~al.}(2013)\citenamefont {Hirano},
  \citenamefont {Takamori}, \citenamefont {Ichioka},\ and\ \citenamefont
  {Machida}}]{Hirano2013}%
  \BibitemOpen
  \bibfield  {author} {\bibinfo {author} {\bibfnamefont {T.}~\bibnamefont
  {Hirano}}, \bibinfo {author} {\bibfnamefont {K.}~\bibnamefont {Takamori}},
  \bibinfo {author} {\bibfnamefont {M.}~\bibnamefont {Ichioka}},\ and\ \bibinfo
  {author} {\bibfnamefont {K.}~\bibnamefont {Machida}},\ }\bibfield  {title}
  {\bibinfo {title} {Rotation of triangular vortex lattice in the two-band
  superconductor {MgB}$_2$},\ }\href {https://doi.org/10.7566/JPSJ.82.063708}
  {\bibfield  {journal} {\bibinfo  {journal} {J. Phys. Soc. Jpn.}\ }\textbf
  {\bibinfo {volume} {82}},\ \bibinfo {pages} {063708} (\bibinfo {year}
  {2013})},\ \Eprint
  {https://arxiv.org/abs/https://doi.org/10.7566/JPSJ.82.063708}
  {https://doi.org/10.7566/JPSJ.82.063708} \BibitemShut {NoStop}%
\bibitem [{\citenamefont {Leishman}\ \emph {et~al.}(2021)\citenamefont
  {Leishman}, \citenamefont {Sokolova}, \citenamefont {Bleuel}, \citenamefont
  {Zhigadlo},\ and\ \citenamefont {Eskildsen}}]{Leishman2021}%
  \BibitemOpen
  \bibfield  {author} {\bibinfo {author} {\bibfnamefont {A.~W.~D.}\
  \bibnamefont {Leishman}}, \bibinfo {author} {\bibfnamefont {A.}~\bibnamefont
  {Sokolova}}, \bibinfo {author} {\bibfnamefont {M.}~\bibnamefont {Bleuel}},
  \bibinfo {author} {\bibfnamefont {N.~D.}\ \bibnamefont {Zhigadlo}},\ and\
  \bibinfo {author} {\bibfnamefont {M.~R.}\ \bibnamefont {Eskildsen}},\
  }\bibfield  {title} {\bibinfo {title} {Field angle dependent vortex lattice
  phase diagram in {MgB}$_{2}$},\ }\href
  {https://doi.org/10.1103/PhysRevB.103.094516} {\bibfield  {journal} {\bibinfo
   {journal} {Phys. Rev. B}\ }\textbf {\bibinfo {volume} {103}},\ \bibinfo
  {pages} {094516} (\bibinfo {year} {2021})}\BibitemShut {NoStop}%
\bibitem [{\citenamefont {Agterberg}(1998{\natexlab{a}})}]{Agterberg1998PRL}%
  \BibitemOpen
  \bibfield  {author} {\bibinfo {author} {\bibfnamefont {D.~F.}\ \bibnamefont
  {Agterberg}},\ }\bibfield  {title} {\bibinfo {title} {Vortex lattice
  structures of {Sr}$_{2}${RuO}$_{4}$},\ }\href
  {https://doi.org/10.1103/PhysRevLett.80.5184} {\bibfield  {journal} {\bibinfo
   {journal} {Phys. Rev. Lett.}\ }\textbf {\bibinfo {volume} {80}},\ \bibinfo
  {pages} {5184} (\bibinfo {year} {1998}{\natexlab{a}})}\BibitemShut {NoStop}%
\bibitem [{\citenamefont {Agterberg}(1998{\natexlab{b}})}]{Agterberg1998PRB}%
  \BibitemOpen
  \bibfield  {author} {\bibinfo {author} {\bibfnamefont {D.~F.}\ \bibnamefont
  {Agterberg}},\ }\bibfield  {title} {\bibinfo {title} {Square vortex lattices
  for two-component superconducting order parameters},\ }\href
  {https://doi.org/10.1103/PhysRevB.58.14484} {\bibfield  {journal} {\bibinfo
  {journal} {Phys. Rev. B}\ }\textbf {\bibinfo {volume} {58}},\ \bibinfo
  {pages} {14484} (\bibinfo {year} {1998}{\natexlab{b}})}\BibitemShut {NoStop}%
\bibitem [{\citenamefont {Riseman}\ \emph {et~al.}(1998)\citenamefont
  {Riseman}, \citenamefont {Kealey}, \citenamefont {Forgan}, \citenamefont
  {Mackenzie}, \citenamefont {Galvin}, \citenamefont {Tyler}, \citenamefont
  {Lee}, \citenamefont {Ager}, \citenamefont {Paul}, \citenamefont {Aegerter},
  \citenamefont {Cubitt}, \citenamefont {Mao}, \citenamefont {Akima},\ and\
  \citenamefont {Maeno}}]{Riseman1998}%
  \BibitemOpen
  \bibfield  {author} {\bibinfo {author} {\bibfnamefont {T.~M.}\ \bibnamefont
  {Riseman}}, \bibinfo {author} {\bibfnamefont {P.~G.}\ \bibnamefont {Kealey}},
  \bibinfo {author} {\bibfnamefont {E.~M.}\ \bibnamefont {Forgan}}, \bibinfo
  {author} {\bibfnamefont {A.~P.}\ \bibnamefont {Mackenzie}}, \bibinfo {author}
  {\bibfnamefont {L.~M.}\ \bibnamefont {Galvin}}, \bibinfo {author}
  {\bibfnamefont {A.~W.}\ \bibnamefont {Tyler}}, \bibinfo {author}
  {\bibfnamefont {S.~L.}\ \bibnamefont {Lee}}, \bibinfo {author} {\bibfnamefont
  {C.}~\bibnamefont {Ager}}, \bibinfo {author} {\bibfnamefont {D.~M.}\
  \bibnamefont {Paul}}, \bibinfo {author} {\bibfnamefont {C.~M.}\ \bibnamefont
  {Aegerter}}, \bibinfo {author} {\bibfnamefont {R.}~\bibnamefont {Cubitt}},
  \bibinfo {author} {\bibfnamefont {Z.~Q.}\ \bibnamefont {Mao}}, \bibinfo
  {author} {\bibfnamefont {T.}~\bibnamefont {Akima}},\ and\ \bibinfo {author}
  {\bibfnamefont {Y.}~\bibnamefont {Maeno}},\ }\bibfield  {title} {\bibinfo
  {title} {Observation of a square flux-line lattice in the unconventional
  superconductor {Sr}$_2${RuO}$_4$},\ }\href {https://doi.org/10.1038/24335}
  {\bibfield  {journal} {\bibinfo  {journal} {Nature}\ }\textbf {\bibinfo
  {volume} {396}},\ \bibinfo {pages} {242} (\bibinfo {year}
  {1998})}\BibitemShut {NoStop}%
\bibitem [{\citenamefont {Huxley}\ \emph {et~al.}(2000)\citenamefont {Huxley},
  \citenamefont {Rodi{\`e}re}, \citenamefont {Paul}, \citenamefont {van Dijk},
  \citenamefont {Cubitt},\ and\ \citenamefont {Flouquet}}]{Huxley2000}%
  \BibitemOpen
  \bibfield  {author} {\bibinfo {author} {\bibfnamefont {A.}~\bibnamefont
  {Huxley}}, \bibinfo {author} {\bibfnamefont {P.}~\bibnamefont {Rodi{\`e}re}},
  \bibinfo {author} {\bibfnamefont {D.~M.}\ \bibnamefont {Paul}}, \bibinfo
  {author} {\bibfnamefont {N.}~\bibnamefont {van Dijk}}, \bibinfo {author}
  {\bibfnamefont {R.}~\bibnamefont {Cubitt}},\ and\ \bibinfo {author}
  {\bibfnamefont {J.}~\bibnamefont {Flouquet}},\ }\bibfield  {title} {\bibinfo
  {title} {Realignment of the flux-line lattice by a change in the symmetry of
  superconductivity in {UPt}$_3$},\ }\href {https://doi.org/10.1038/35018020}
  {\bibfield  {journal} {\bibinfo  {journal} {Nature}\ }\textbf {\bibinfo
  {volume} {406}},\ \bibinfo {pages} {160} (\bibinfo {year}
  {2000})}\BibitemShut {NoStop}%
\bibitem [{\citenamefont {Brown}\ \emph {et~al.}(2004)\citenamefont {Brown},
  \citenamefont {Charalambous}, \citenamefont {Jones}, \citenamefont {Forgan},
  \citenamefont {Kealey}, \citenamefont {Erb},\ and\ \citenamefont
  {Kohlbrecher}}]{Brown2004}%
  \BibitemOpen
  \bibfield  {author} {\bibinfo {author} {\bibfnamefont {S.~P.}\ \bibnamefont
  {Brown}}, \bibinfo {author} {\bibfnamefont {D.}~\bibnamefont {Charalambous}},
  \bibinfo {author} {\bibfnamefont {E.~C.}\ \bibnamefont {Jones}}, \bibinfo
  {author} {\bibfnamefont {E.~M.}\ \bibnamefont {Forgan}}, \bibinfo {author}
  {\bibfnamefont {P.~G.}\ \bibnamefont {Kealey}}, \bibinfo {author}
  {\bibfnamefont {A.}~\bibnamefont {Erb}},\ and\ \bibinfo {author}
  {\bibfnamefont {J.}~\bibnamefont {Kohlbrecher}},\ }\bibfield  {title}
  {\bibinfo {title} {Triangular to square flux lattice phase transition in
  {Y}{Ba}$_{2}${Cu}$_{3}${O}$_{7}$},\ }\href
  {https://doi.org/10.1103/PhysRevLett.92.067004} {\bibfield  {journal}
  {\bibinfo  {journal} {Phys. Rev. Lett.}\ }\textbf {\bibinfo {volume} {92}},\
  \bibinfo {pages} {067004} (\bibinfo {year} {2004})}\BibitemShut {NoStop}%
\bibitem [{\citenamefont {Bianchi}\ \emph {et~al.}(2008)\citenamefont
  {Bianchi}, \citenamefont {Kenzelmann}, \citenamefont {DeBeer-Schmitt},
  \citenamefont {White}, \citenamefont {Forgan}, \citenamefont {Mesot},
  \citenamefont {Zolliker}, \citenamefont {Kohlbrecher}, \citenamefont
  {Movshovich}, \citenamefont {Bauer}, \citenamefont {Sarrao}, \citenamefont
  {Fisk}, \citenamefont {Petrović},\ and\ \citenamefont
  {Eskildsen}}]{Bianchi2008}%
  \BibitemOpen
  \bibfield  {author} {\bibinfo {author} {\bibfnamefont {A.~D.}\ \bibnamefont
  {Bianchi}}, \bibinfo {author} {\bibfnamefont {M.}~\bibnamefont {Kenzelmann}},
  \bibinfo {author} {\bibfnamefont {L.}~\bibnamefont {DeBeer-Schmitt}},
  \bibinfo {author} {\bibfnamefont {J.~S.}\ \bibnamefont {White}}, \bibinfo
  {author} {\bibfnamefont {E.~M.}\ \bibnamefont {Forgan}}, \bibinfo {author}
  {\bibfnamefont {J.}~\bibnamefont {Mesot}}, \bibinfo {author} {\bibfnamefont
  {M.}~\bibnamefont {Zolliker}}, \bibinfo {author} {\bibfnamefont
  {J.}~\bibnamefont {Kohlbrecher}}, \bibinfo {author} {\bibfnamefont
  {R.}~\bibnamefont {Movshovich}}, \bibinfo {author} {\bibfnamefont {E.~D.}\
  \bibnamefont {Bauer}}, \bibinfo {author} {\bibfnamefont {J.~L.}\ \bibnamefont
  {Sarrao}}, \bibinfo {author} {\bibfnamefont {Z.}~\bibnamefont {Fisk}},
  \bibinfo {author} {\bibfnamefont {C.}~\bibnamefont {Petrović}},\ and\
  \bibinfo {author} {\bibfnamefont {M.~R.}\ \bibnamefont {Eskildsen}},\
  }\bibfield  {title} {\bibinfo {title} {Superconducting vortices in
  {CeCoIn}$_5$: Toward the {P}auli-limiting field},\ }\href
  {https://doi.org/10.1126/science.1150600} {\bibfield  {journal} {\bibinfo
  {journal} {Science}\ }\textbf {\bibinfo {volume} {319}},\ \bibinfo {pages}
  {177} (\bibinfo {year} {2008})}\BibitemShut {NoStop}%
\bibitem [{\citenamefont {Avers}\ \emph {et~al.}(2020)\citenamefont {Avers},
  \citenamefont {Gannon}, \citenamefont {Kuhn}, \citenamefont {Halperin},
  \citenamefont {Sauls}, \citenamefont {DeBeer-Schmitt}, \citenamefont
  {Dewhurst}, \citenamefont {Gavilano}, \citenamefont {Nagy}, \citenamefont
  {Gasser},\ and\ \citenamefont {Eskildsen}}]{Avers2020}%
  \BibitemOpen
  \bibfield  {author} {\bibinfo {author} {\bibfnamefont {K.~E.}\ \bibnamefont
  {Avers}}, \bibinfo {author} {\bibfnamefont {W.~J.}\ \bibnamefont {Gannon}},
  \bibinfo {author} {\bibfnamefont {S.~J.}\ \bibnamefont {Kuhn}}, \bibinfo
  {author} {\bibfnamefont {W.~P.}\ \bibnamefont {Halperin}}, \bibinfo {author}
  {\bibfnamefont {J.~A.}\ \bibnamefont {Sauls}}, \bibinfo {author}
  {\bibfnamefont {L.}~\bibnamefont {DeBeer-Schmitt}}, \bibinfo {author}
  {\bibfnamefont {C.~D.}\ \bibnamefont {Dewhurst}}, \bibinfo {author}
  {\bibfnamefont {J.}~\bibnamefont {Gavilano}}, \bibinfo {author}
  {\bibfnamefont {G.}~\bibnamefont {Nagy}}, \bibinfo {author} {\bibfnamefont
  {U.}~\bibnamefont {Gasser}},\ and\ \bibinfo {author} {\bibfnamefont {M.~R.}\
  \bibnamefont {Eskildsen}},\ }\bibfield  {title} {\bibinfo {title} {Broken
  time-reversal symmetry in the topological superconductor {UPt}$_3$},\ }\href
  {https://doi.org/10.1038/s41567-020-0822-z} {\bibfield  {journal} {\bibinfo
  {journal} {Nature Phys.}\ }\textbf {\bibinfo {volume} {16}},\ \bibinfo
  {pages} {531} (\bibinfo {year} {2020})}\BibitemShut {NoStop}%
\bibitem [{\citenamefont {Avers}\ \emph {et~al.}(2022)\citenamefont {Avers},
  \citenamefont {Gannon}, \citenamefont {Leishman}, \citenamefont
  {DeBeer-Schmitt}, \citenamefont {Halperin},\ and\ \citenamefont
  {Eskildsen}}]{Avers2022}%
  \BibitemOpen
  \bibfield  {author} {\bibinfo {author} {\bibfnamefont {K.~E.}\ \bibnamefont
  {Avers}}, \bibinfo {author} {\bibfnamefont {W.~J.}\ \bibnamefont {Gannon}},
  \bibinfo {author} {\bibfnamefont {A.~W.~D.}\ \bibnamefont {Leishman}},
  \bibinfo {author} {\bibfnamefont {L.}~\bibnamefont {DeBeer-Schmitt}},
  \bibinfo {author} {\bibfnamefont {W.~P.}\ \bibnamefont {Halperin}},\ and\
  \bibinfo {author} {\bibfnamefont {M.~R.}\ \bibnamefont {Eskildsen}},\
  }\bibfield  {title} {\bibinfo {title} {Effects of the order parameter
  anisotropy on the vortex lattice in {UPt}$_3$},\ }\bibfield  {journal}
  {\bibinfo  {journal} {Front. Electron. Mater.}\ }\textbf {\bibinfo {volume}
  {2}},\ \href {https://doi.org/10.3389/femat.2022.878308}
  {10.3389/femat.2022.878308} (\bibinfo {year} {2022})\BibitemShut {NoStop}%
\bibitem [{\citenamefont {Hu}\ \emph {et~al.}(2012)\citenamefont {Hu},
  \citenamefont {Xiao}, \citenamefont {Sayles}, \citenamefont {Dzero},
  \citenamefont {Maple},\ and\ \citenamefont {Almasan}}]{Hu2012}%
  \BibitemOpen
  \bibfield  {author} {\bibinfo {author} {\bibfnamefont {T.}~\bibnamefont
  {Hu}}, \bibinfo {author} {\bibfnamefont {H.}~\bibnamefont {Xiao}}, \bibinfo
  {author} {\bibfnamefont {T.~A.}\ \bibnamefont {Sayles}}, \bibinfo {author}
  {\bibfnamefont {M.}~\bibnamefont {Dzero}}, \bibinfo {author} {\bibfnamefont
  {M.~B.}\ \bibnamefont {Maple}},\ and\ \bibinfo {author} {\bibfnamefont
  {C.~C.}\ \bibnamefont {Almasan}},\ }\bibfield  {title} {\bibinfo {title}
  {Strong magnetic fluctuations in a superconducting state of {CeCoIn}$_{5}$},\
  }\href {https://doi.org/10.1103/PhysRevLett.108.056401} {\bibfield  {journal}
  {\bibinfo  {journal} {Phys. Rev. Lett.}\ }\textbf {\bibinfo {volume} {108}},\
  \bibinfo {pages} {056401} (\bibinfo {year} {2012})}\BibitemShut {NoStop}%
\bibitem [{\citenamefont {Jang}\ \emph {et~al.}(2016)\citenamefont {Jang},
  \citenamefont {Pedrero}, \citenamefont {Pham}, \citenamefont {Fisk},\ and\
  \citenamefont {Brando}}]{Jang2016}%
  \BibitemOpen
  \bibfield  {author} {\bibinfo {author} {\bibfnamefont {D.-J.}\ \bibnamefont
  {Jang}}, \bibinfo {author} {\bibfnamefont {L.}~\bibnamefont {Pedrero}},
  \bibinfo {author} {\bibfnamefont {L.~D.}\ \bibnamefont {Pham}}, \bibinfo
  {author} {\bibfnamefont {Z.}~\bibnamefont {Fisk}},\ and\ \bibinfo {author}
  {\bibfnamefont {M.}~\bibnamefont {Brando}},\ }\bibfield  {title} {\bibinfo
  {title} {Strong pinning of vortices by antiferromagnetic domain boundaries in
  {CeCo}({In}$_{1-x}${Cd}$_x$)$_5$},\ }\href
  {https://doi.org/10.1088/1367-2630/18/9/093031} {\bibfield  {journal}
  {\bibinfo  {journal} {New J. Phys.}\ }\textbf {\bibinfo {volume} {18}},\
  \bibinfo {pages} {093031} (\bibinfo {year} {2016})}\BibitemShut {NoStop}%
\bibitem [{\citenamefont {Wulferding}\ \emph {et~al.}(2020)\citenamefont
  {Wulferding}, \citenamefont {Kim}, \citenamefont {Kim}, \citenamefont {Yang},
  \citenamefont {Bauer}, \citenamefont {Ronning}, \citenamefont {Movshovich},\
  and\ \citenamefont {Kim}}]{Wulferding2020}%
  \BibitemOpen
  \bibfield  {author} {\bibinfo {author} {\bibfnamefont {D.}~\bibnamefont
  {Wulferding}}, \bibinfo {author} {\bibfnamefont {G.}~\bibnamefont {Kim}},
  \bibinfo {author} {\bibfnamefont {H.}~\bibnamefont {Kim}}, \bibinfo {author}
  {\bibfnamefont {I.}~\bibnamefont {Yang}}, \bibinfo {author} {\bibfnamefont
  {E.~D.}\ \bibnamefont {Bauer}}, \bibinfo {author} {\bibfnamefont
  {F.}~\bibnamefont {Ronning}}, \bibinfo {author} {\bibfnamefont
  {R.}~\bibnamefont {Movshovich}},\ and\ \bibinfo {author} {\bibfnamefont
  {J.}~\bibnamefont {Kim}},\ }\bibfield  {title} {\bibinfo {title} {Local
  characterization of a heavy-fermion superconductor via sub-{K}elvin magnetic
  force microscopy},\ }\href {https://doi.org/10.1063/5.0028517} {\bibfield
  {journal} {\bibinfo  {journal} {Appl. Phys. Lett.}\ }\textbf {\bibinfo
  {volume} {117}},\ \bibinfo {pages} {252601} (\bibinfo {year}
  {2020})}\BibitemShut {NoStop}%
\bibitem [{\citenamefont {Eskildsen}\ \emph {et~al.}(2003)\citenamefont
  {Eskildsen}, \citenamefont {Dewhurst}, \citenamefont {Hoogenboom},
  \citenamefont {Petrovic},\ and\ \citenamefont {Canfield}}]{Eskildsen2003}%
  \BibitemOpen
  \bibfield  {author} {\bibinfo {author} {\bibfnamefont {M.~R.}\ \bibnamefont
  {Eskildsen}}, \bibinfo {author} {\bibfnamefont {C.~D.}\ \bibnamefont
  {Dewhurst}}, \bibinfo {author} {\bibfnamefont {B.~W.}\ \bibnamefont
  {Hoogenboom}}, \bibinfo {author} {\bibfnamefont {C.}~\bibnamefont
  {Petrovic}},\ and\ \bibinfo {author} {\bibfnamefont {P.~C.}\ \bibnamefont
  {Canfield}},\ }\bibfield  {title} {\bibinfo {title} {Hexagonal and square
  flux line lattices in {CeCoIn}$_{5}$},\ }\href
  {https://doi.org/10.1103/PhysRevLett.90.187001} {\bibfield  {journal}
  {\bibinfo  {journal} {Phys. Rev. Lett.}\ }\textbf {\bibinfo {volume} {90}},\
  \bibinfo {pages} {187001} (\bibinfo {year} {2003})}\BibitemShut {NoStop}%
\bibitem [{\citenamefont {Ohira-Kawamura}\ \emph {et~al.}(2008)\citenamefont
  {Ohira-Kawamura}, \citenamefont {Shishido}, \citenamefont {Kawano-Furukawa},
  \citenamefont {Lake}, \citenamefont {Wiedenmann}, \citenamefont {Kiefer},
  \citenamefont {Shibauchi},\ and\ \citenamefont
  {Matsuda}}]{Ohira-Kawamura2008}%
  \BibitemOpen
  \bibfield  {author} {\bibinfo {author} {\bibfnamefont {S.}~\bibnamefont
  {Ohira-Kawamura}}, \bibinfo {author} {\bibfnamefont {H.}~\bibnamefont
  {Shishido}}, \bibinfo {author} {\bibfnamefont {H.}~\bibnamefont
  {Kawano-Furukawa}}, \bibinfo {author} {\bibfnamefont {B.}~\bibnamefont
  {Lake}}, \bibinfo {author} {\bibfnamefont {A.}~\bibnamefont {Wiedenmann}},
  \bibinfo {author} {\bibfnamefont {K.}~\bibnamefont {Kiefer}}, \bibinfo
  {author} {\bibfnamefont {T.}~\bibnamefont {Shibauchi}},\ and\ \bibinfo
  {author} {\bibfnamefont {Y.}~\bibnamefont {Matsuda}},\ }\bibfield  {title}
  {\bibinfo {title} {Anomalous flux line lattice in {CeCoIn}$_5$},\ }\href
  {https://doi.org/10.1143/JPSJ.77.023702} {\bibfield  {journal} {\bibinfo
  {journal} {J. Phys. Soc. Jpn.}\ }\textbf {\bibinfo {volume} {77}},\ \bibinfo
  {pages} {023702} (\bibinfo {year} {2008})}\BibitemShut {NoStop}%
\bibitem [{\citenamefont {White}\ \emph {et~al.}(2010)\citenamefont {White},
  \citenamefont {Das}, \citenamefont {Eskildsen}, \citenamefont
  {DeBeer-Schmitt}, \citenamefont {Forgan}, \citenamefont {Bianchi},
  \citenamefont {Kenzelmann}, \citenamefont {Zolliker}, \citenamefont {Gerber},
  \citenamefont {Gavilano}, \citenamefont {Mesot}, \citenamefont {Movshovich},
  \citenamefont {Bauer}, \citenamefont {Sarrao},\ and\ \citenamefont
  {Petrovic}}]{White2010}%
  \BibitemOpen
  \bibfield  {author} {\bibinfo {author} {\bibfnamefont {J.~S.}\ \bibnamefont
  {White}}, \bibinfo {author} {\bibfnamefont {P.}~\bibnamefont {Das}}, \bibinfo
  {author} {\bibfnamefont {M.~R.}\ \bibnamefont {Eskildsen}}, \bibinfo {author}
  {\bibfnamefont {L.}~\bibnamefont {DeBeer-Schmitt}}, \bibinfo {author}
  {\bibfnamefont {E.~M.}\ \bibnamefont {Forgan}}, \bibinfo {author}
  {\bibfnamefont {A.~D.}\ \bibnamefont {Bianchi}}, \bibinfo {author}
  {\bibfnamefont {M.}~\bibnamefont {Kenzelmann}}, \bibinfo {author}
  {\bibfnamefont {M.}~\bibnamefont {Zolliker}}, \bibinfo {author}
  {\bibfnamefont {S.}~\bibnamefont {Gerber}}, \bibinfo {author} {\bibfnamefont
  {J.~L.}\ \bibnamefont {Gavilano}}, \bibinfo {author} {\bibfnamefont
  {J.}~\bibnamefont {Mesot}}, \bibinfo {author} {\bibfnamefont
  {R.}~\bibnamefont {Movshovich}}, \bibinfo {author} {\bibfnamefont {E.~D.}\
  \bibnamefont {Bauer}}, \bibinfo {author} {\bibfnamefont {J.~L.}\ \bibnamefont
  {Sarrao}},\ and\ \bibinfo {author} {\bibfnamefont {C.}~\bibnamefont
  {Petrovic}},\ }\bibfield  {title} {\bibinfo {title} {Observations of {P}auli
  paramagnetic effects on the flux line lattice in {CeCoIn}$_5$},\ }\href
  {https://doi.org/10.1088/1367-2630/12/2/023026} {\bibfield  {journal}
  {\bibinfo  {journal} {New J. Phys.}\ }\textbf {\bibinfo {volume} {12}},\
  \bibinfo {pages} {023026} (\bibinfo {year} {2010})}\BibitemShut {NoStop}%
\bibitem [{\citenamefont {DeBeer-Schmitt}\ \emph {et~al.}(2006)\citenamefont
  {DeBeer-Schmitt}, \citenamefont {Dewhurst}, \citenamefont {Hoogenboom},
  \citenamefont {Petrovic},\ and\ \citenamefont
  {Eskildsen}}]{DeBeer-Schmitt2006}%
  \BibitemOpen
  \bibfield  {author} {\bibinfo {author} {\bibfnamefont {L.}~\bibnamefont
  {DeBeer-Schmitt}}, \bibinfo {author} {\bibfnamefont {C.~D.}\ \bibnamefont
  {Dewhurst}}, \bibinfo {author} {\bibfnamefont {B.~W.}\ \bibnamefont
  {Hoogenboom}}, \bibinfo {author} {\bibfnamefont {C.}~\bibnamefont
  {Petrovic}},\ and\ \bibinfo {author} {\bibfnamefont {M.~R.}\ \bibnamefont
  {Eskildsen}},\ }\bibfield  {title} {\bibinfo {title} {Field dependent
  coherence length in the superclean, high-$\ensuremath{\kappa}$ superconductor
  {CeCoIn}$_{5}$},\ }\href {https://doi.org/10.1103/PhysRevLett.97.127001}
  {\bibfield  {journal} {\bibinfo  {journal} {Phys. Rev. Lett.}\ }\textbf
  {\bibinfo {volume} {97}},\ \bibinfo {pages} {127001} (\bibinfo {year}
  {2006})}\BibitemShut {NoStop}%
\bibitem [{\citenamefont {Das}\ \emph {et~al.}(2012{\natexlab{b}})\citenamefont
  {Das}, \citenamefont {White}, \citenamefont {Holmes}, \citenamefont {Gerber},
  \citenamefont {Forgan}, \citenamefont {Bianchi}, \citenamefont {Kenzelmann},
  \citenamefont {Zolliker}, \citenamefont {Gavilano}, \citenamefont {Bauer},
  \citenamefont {Sarrao}, \citenamefont {Petrovic},\ and\ \citenamefont
  {Eskildsen}}]{Das2012CeCoIn5}%
  \BibitemOpen
  \bibfield  {author} {\bibinfo {author} {\bibfnamefont {P.}~\bibnamefont
  {Das}}, \bibinfo {author} {\bibfnamefont {J.~S.}\ \bibnamefont {White}},
  \bibinfo {author} {\bibfnamefont {A.~T.}\ \bibnamefont {Holmes}}, \bibinfo
  {author} {\bibfnamefont {S.}~\bibnamefont {Gerber}}, \bibinfo {author}
  {\bibfnamefont {E.~M.}\ \bibnamefont {Forgan}}, \bibinfo {author}
  {\bibfnamefont {A.~D.}\ \bibnamefont {Bianchi}}, \bibinfo {author}
  {\bibfnamefont {M.}~\bibnamefont {Kenzelmann}}, \bibinfo {author}
  {\bibfnamefont {M.}~\bibnamefont {Zolliker}}, \bibinfo {author}
  {\bibfnamefont {J.~L.}\ \bibnamefont {Gavilano}}, \bibinfo {author}
  {\bibfnamefont {E.~D.}\ \bibnamefont {Bauer}}, \bibinfo {author}
  {\bibfnamefont {J.~L.}\ \bibnamefont {Sarrao}}, \bibinfo {author}
  {\bibfnamefont {C.}~\bibnamefont {Petrovic}},\ and\ \bibinfo {author}
  {\bibfnamefont {M.~R.}\ \bibnamefont {Eskildsen}},\ }\bibfield  {title}
  {\bibinfo {title} {Vortex lattice studies in {CeCoIn}$_{5}$ with
  $h\ensuremath{\perp}c$},\ }\href
  {https://doi.org/10.1103/PhysRevLett.108.087002} {\bibfield  {journal}
  {\bibinfo  {journal} {Phys. Rev. Lett.}\ }\textbf {\bibinfo {volume} {108}},\
  \bibinfo {pages} {087002} (\bibinfo {year} {2012}{\natexlab{b}})}\BibitemShut
  {NoStop}%
\bibitem [{\citenamefont {Tayama}\ \emph {et~al.}(2002)\citenamefont {Tayama},
  \citenamefont {Harita}, \citenamefont {Sakakibara}, \citenamefont {Haga},
  \citenamefont {Shishido}, \citenamefont {Settai},\ and\ \citenamefont
  {Onuki}}]{Tayama2002}%
  \BibitemOpen
  \bibfield  {author} {\bibinfo {author} {\bibfnamefont {T.}~\bibnamefont
  {Tayama}}, \bibinfo {author} {\bibfnamefont {A.}~\bibnamefont {Harita}},
  \bibinfo {author} {\bibfnamefont {T.}~\bibnamefont {Sakakibara}}, \bibinfo
  {author} {\bibfnamefont {Y.}~\bibnamefont {Haga}}, \bibinfo {author}
  {\bibfnamefont {H.}~\bibnamefont {Shishido}}, \bibinfo {author}
  {\bibfnamefont {R.}~\bibnamefont {Settai}},\ and\ \bibinfo {author}
  {\bibfnamefont {Y.}~\bibnamefont {Onuki}},\ }\bibfield  {title} {\bibinfo
  {title} {Unconventional heavy-fermion superconductor {CeCoIn}$_{5}:$ dc
  magnetization study at temperatures down to 50 m{K}},\ }\href
  {https://doi.org/10.1103/PhysRevB.65.180504} {\bibfield  {journal} {\bibinfo
  {journal} {Phys. Rev. B}\ }\textbf {\bibinfo {volume} {65}},\ \bibinfo
  {pages} {180504} (\bibinfo {year} {2002})}\BibitemShut {NoStop}%
\bibitem [{\citenamefont {Prommapan}\ \emph {et~al.}(2011)\citenamefont
  {Prommapan}, \citenamefont {Tanatar}, \citenamefont {Lee}, \citenamefont
  {Khim}, \citenamefont {Kim},\ and\ \citenamefont {Prozorov}}]{Prommapan2011}%
  \BibitemOpen
  \bibfield  {author} {\bibinfo {author} {\bibfnamefont {P.}~\bibnamefont
  {Prommapan}}, \bibinfo {author} {\bibfnamefont {M.~A.}\ \bibnamefont
  {Tanatar}}, \bibinfo {author} {\bibfnamefont {B.}~\bibnamefont {Lee}},
  \bibinfo {author} {\bibfnamefont {S.}~\bibnamefont {Khim}}, \bibinfo {author}
  {\bibfnamefont {K.~H.}\ \bibnamefont {Kim}},\ and\ \bibinfo {author}
  {\bibfnamefont {R.}~\bibnamefont {Prozorov}},\ }\bibfield  {title} {\bibinfo
  {title} {Magnetic-field-dependent pinning potential in {LiFeAs}
  superconductor from its campbell penetration depth},\ }\href
  {https://doi.org/10.1103/PhysRevB.84.060509} {\bibfield  {journal} {\bibinfo
  {journal} {Phys. Rev. B}\ }\textbf {\bibinfo {volume} {84}},\ \bibinfo
  {pages} {060509} (\bibinfo {year} {2011})}\BibitemShut {NoStop}%
\bibitem [{\citenamefont {Kim}\ \emph {et~al.}(2013)\citenamefont {Kim},
  \citenamefont {Sung}, \citenamefont {Cho}, \citenamefont {Tanatar},\ and\
  \citenamefont {Prozorov}}]{Kim2013}%
  \BibitemOpen
  \bibfield  {author} {\bibinfo {author} {\bibfnamefont {H.}~\bibnamefont
  {Kim}}, \bibinfo {author} {\bibfnamefont {N.~H.}\ \bibnamefont {Sung}},
  \bibinfo {author} {\bibfnamefont {B.~K.}\ \bibnamefont {Cho}}, \bibinfo
  {author} {\bibfnamefont {M.~A.}\ \bibnamefont {Tanatar}},\ and\ \bibinfo
  {author} {\bibfnamefont {R.}~\bibnamefont {Prozorov}},\ }\bibfield  {title}
  {\bibinfo {title} {Magnetic penetration depth in single crystals of
  {SrPd}$_{2}${Ge}$_{2}$ superconductor},\ }\href
  {https://doi.org/10.1103/PhysRevB.87.094515} {\bibfield  {journal} {\bibinfo
  {journal} {Phys. Rev. B}\ }\textbf {\bibinfo {volume} {87}},\ \bibinfo
  {pages} {094515} (\bibinfo {year} {2013})}\BibitemShut {NoStop}%
\bibitem [{\citenamefont {Kim}\ \emph {et~al.}(2021)\citenamefont {Kim},
  \citenamefont {Tanatar}, \citenamefont {Hodovanets}, \citenamefont {Wang},
  \citenamefont {Paglione},\ and\ \citenamefont {Prozorov}}]{Kim2021}%
  \BibitemOpen
  \bibfield  {author} {\bibinfo {author} {\bibfnamefont {H.}~\bibnamefont
  {Kim}}, \bibinfo {author} {\bibfnamefont {M.~A.}\ \bibnamefont {Tanatar}},
  \bibinfo {author} {\bibfnamefont {H.}~\bibnamefont {Hodovanets}}, \bibinfo
  {author} {\bibfnamefont {K.}~\bibnamefont {Wang}}, \bibinfo {author}
  {\bibfnamefont {J.}~\bibnamefont {Paglione}},\ and\ \bibinfo {author}
  {\bibfnamefont {R.}~\bibnamefont {Prozorov}},\ }\bibfield  {title} {\bibinfo
  {title} {Campbell penetration depth in low carrier density superconductor
  {YPtBi}},\ }\href {https://doi.org/10.1103/PhysRevB.104.014510} {\bibfield
  {journal} {\bibinfo  {journal} {Phys. Rev. B}\ }\textbf {\bibinfo {volume}
  {104}},\ \bibinfo {pages} {014510} (\bibinfo {year} {2021})}\BibitemShut
  {NoStop}%
\bibitem [{\citenamefont {Campbell}(1969)}]{Campbell1969}%
  \BibitemOpen
  \bibfield  {author} {\bibinfo {author} {\bibfnamefont {A.~M.}\ \bibnamefont
  {Campbell}},\ }\bibfield  {title} {\bibinfo {title} {The response of pinned
  flux vortices to low-frequency fields},\ }\href
  {http://stacks.iop.org/0022-3719/2/i=8/a=318} {\bibfield  {journal} {\bibinfo
   {journal} {J. Phys. C}\ }\textbf {\bibinfo {volume} {2}},\ \bibinfo {pages}
  {1492} (\bibinfo {year} {1969})}\BibitemShut {NoStop}%
\bibitem [{\citenamefont {Campbell}(1971)}]{Campbell1971}%
  \BibitemOpen
  \bibfield  {author} {\bibinfo {author} {\bibfnamefont {A.~M.}\ \bibnamefont
  {Campbell}},\ }\bibfield  {title} {\bibinfo {title} {The interaction distance
  between flux lines and pinning centres},\ }\href
  {http://stacks.iop.org/0022-3719/4/i=18/a=023} {\bibfield  {journal}
  {\bibinfo  {journal} {J. Phys. C}\ }\textbf {\bibinfo {volume} {4}},\
  \bibinfo {pages} {3186} (\bibinfo {year} {1971})}\BibitemShut {NoStop}%
\bibitem [{\citenamefont {Brandt}(1995)}]{Brandt1995}%
  \BibitemOpen
  \bibfield  {author} {\bibinfo {author} {\bibfnamefont {E.~H.}\ \bibnamefont
  {Brandt}},\ }\bibfield  {title} {\bibinfo {title} {The flux-line lattice in
  superconductors},\ }\href {http://stacks.iop.org/0034-4885/58/1465}
  {\bibfield  {journal} {\bibinfo  {journal} {Rep. Prog. Phys.}\ }\textbf
  {\bibinfo {volume} {58}},\ \bibinfo {pages} {1465} (\bibinfo {year}
  {1995})}\BibitemShut {NoStop}%
\bibitem [{\citenamefont {Prozorov}\ \emph {et~al.}(2003)\citenamefont
  {Prozorov}, \citenamefont {Giannetta}, \citenamefont {Kameda}, \citenamefont
  {Tamegai}, \citenamefont {Schlueter},\ and\ \citenamefont
  {Fournier}}]{Prozorov2003e}%
  \BibitemOpen
  \bibfield  {author} {\bibinfo {author} {\bibfnamefont {R.}~\bibnamefont
  {Prozorov}}, \bibinfo {author} {\bibfnamefont {R.~W.}\ \bibnamefont
  {Giannetta}}, \bibinfo {author} {\bibfnamefont {N.}~\bibnamefont {Kameda}},
  \bibinfo {author} {\bibfnamefont {T.}~\bibnamefont {Tamegai}}, \bibinfo
  {author} {\bibfnamefont {J.~A.}\ \bibnamefont {Schlueter}},\ and\ \bibinfo
  {author} {\bibfnamefont {P.}~\bibnamefont {Fournier}},\ }\bibfield  {title}
  {\bibinfo {title} {Campbell penetration depth of a superconductor in the
  critical state},\ }\href {https://doi.org/10.1103/PhysRevB.67.184501}
  {\bibfield  {journal} {\bibinfo  {journal} {Phys. Rev. B}\ }\textbf {\bibinfo
  {volume} {67}},\ \bibinfo {pages} {184501} (\bibinfo {year}
  {2003})}\BibitemShut {NoStop}%
\bibitem [{\citenamefont {Willa}\ \emph
  {et~al.}(2015{\natexlab{a}})\citenamefont {Willa}, \citenamefont
  {Geshkenbein},\ and\ \citenamefont {Blatter}}]{Willa2015prb}%
  \BibitemOpen
  \bibfield  {author} {\bibinfo {author} {\bibfnamefont {R.}~\bibnamefont
  {Willa}}, \bibinfo {author} {\bibfnamefont {V.~B.}\ \bibnamefont
  {Geshkenbein}},\ and\ \bibinfo {author} {\bibfnamefont {G.}~\bibnamefont
  {Blatter}},\ }\bibfield  {title} {\bibinfo {title} {Campbell penetration in
  the critical state of type-{II} superconductors},\ }\href
  {https://doi.org/10.1103/PhysRevB.92.134501} {\bibfield  {journal} {\bibinfo
  {journal} {Phys. Rev. B}\ }\textbf {\bibinfo {volume} {92}},\ \bibinfo
  {pages} {134501} (\bibinfo {year} {2015}{\natexlab{a}})}\BibitemShut
  {NoStop}%
\bibitem [{\citenamefont {Willa}\ \emph
  {et~al.}(2015{\natexlab{b}})\citenamefont {Willa}, \citenamefont
  {Geshkenbein}, \citenamefont {Prozorov},\ and\ \citenamefont
  {Blatter}}]{Willa2015}%
  \BibitemOpen
  \bibfield  {author} {\bibinfo {author} {\bibfnamefont {R.}~\bibnamefont
  {Willa}}, \bibinfo {author} {\bibfnamefont {V.~B.}\ \bibnamefont
  {Geshkenbein}}, \bibinfo {author} {\bibfnamefont {R.}~\bibnamefont
  {Prozorov}},\ and\ \bibinfo {author} {\bibfnamefont {G.}~\bibnamefont
  {Blatter}},\ }\bibfield  {title} {\bibinfo {title} {Campbell response in
  type-{II} superconductors under strong pinning conditions},\ }\href
  {https://doi.org/10.1103/PhysRevLett.115.207001} {\bibfield  {journal}
  {\bibinfo  {journal} {Phys. Rev. Lett.}\ }\textbf {\bibinfo {volume} {115}},\
  \bibinfo {pages} {207001} (\bibinfo {year} {2015}{\natexlab{b}})}\BibitemShut
  {NoStop}%
\bibitem [{\citenamefont {Willa}\ \emph {et~al.}(2016)\citenamefont {Willa},
  \citenamefont {Geshkenbein},\ and\ \citenamefont {Blatter}}]{Willa2016}%
  \BibitemOpen
  \bibfield  {author} {\bibinfo {author} {\bibfnamefont {R.}~\bibnamefont
  {Willa}}, \bibinfo {author} {\bibfnamefont {V.~B.}\ \bibnamefont
  {Geshkenbein}},\ and\ \bibinfo {author} {\bibfnamefont {G.}~\bibnamefont
  {Blatter}},\ }\bibfield  {title} {\bibinfo {title} {Probing the pinning
  landscape in type-{II} superconductors via campbell penetration depth},\
  }\href {https://doi.org/10.1103/PhysRevB.93.064515} {\bibfield  {journal}
  {\bibinfo  {journal} {Phys. Rev. B}\ }\textbf {\bibinfo {volume} {93}},\
  \bibinfo {pages} {064515} (\bibinfo {year} {2016})}\BibitemShut {NoStop}%
\bibitem [{\citenamefont {G{\"o}m{\"o}ry}(1997)}]{Gomory1997}%
  \BibitemOpen
  \bibfield  {author} {\bibinfo {author} {\bibfnamefont {F.}~\bibnamefont
  {G{\"o}m{\"o}ry}},\ }\bibfield  {title} {\bibinfo {title} {Characterization
  of high-temperature superconductors by ac susceptibility measurements},\
  }\href {https://doi.org/10.1088/0953-2048/10/8/001} {\bibfield  {journal}
  {\bibinfo  {journal} {Supercond. Sci. Techn.}\ }\textbf {\bibinfo {volume}
  {10}},\ \bibinfo {pages} {523} (\bibinfo {year} {1997})}\BibitemShut
  {NoStop}%
\bibitem [{\citenamefont {Bean}(1962)}]{Bean1962}%
  \BibitemOpen
  \bibfield  {author} {\bibinfo {author} {\bibfnamefont {C.~P.}\ \bibnamefont
  {Bean}},\ }\bibfield  {title} {\bibinfo {title} {Magnetization of hard
  superconductors},\ }\href {https://doi.org/10.1103/PhysRevLett.8.250}
  {\bibfield  {journal} {\bibinfo  {journal} {Phys. Rev. Lett.}\ }\textbf
  {\bibinfo {volume} {8}},\ \bibinfo {pages} {250} (\bibinfo {year}
  {1962})}\BibitemShut {NoStop}%
\bibitem [{\citenamefont {Bean}(1964)}]{Bean1964}%
  \BibitemOpen
  \bibfield  {author} {\bibinfo {author} {\bibfnamefont {C.~P.}\ \bibnamefont
  {Bean}},\ }\bibfield  {title} {\bibinfo {title} {Magnetization of high-field
  superconductors},\ }\href {https://doi.org/10.1103/RevModPhys.36.31}
  {\bibfield  {journal} {\bibinfo  {journal} {Rev. Mod. Phys.}\ }\textbf
  {\bibinfo {volume} {36}},\ \bibinfo {pages} {31} (\bibinfo {year}
  {1964})}\BibitemShut {NoStop}%
\bibitem [{\citenamefont {Bardeen}(1958)}]{Bardeen1958}%
  \BibitemOpen
  \bibfield  {author} {\bibinfo {author} {\bibfnamefont {J.}~\bibnamefont
  {Bardeen}},\ }\bibfield  {title} {\bibinfo {title} {Two-fluid model of
  superconductivity},\ }\href {https://doi.org/10.1103/PhysRevLett.1.399}
  {\bibfield  {journal} {\bibinfo  {journal} {Phys. Rev. Lett.}\ }\textbf
  {\bibinfo {volume} {1}},\ \bibinfo {pages} {399} (\bibinfo {year}
  {1958})}\BibitemShut {NoStop}%
\bibitem [{\citenamefont {Gor'kov}(1959)}]{Gorkov1959}%
  \BibitemOpen
  \bibfield  {author} {\bibinfo {author} {\bibfnamefont {L.~P.}\ \bibnamefont
  {Gor'kov}},\ }\bibfield  {title} {\bibinfo {title} {Microscopic derivation of
  the {G}inzburg--{L}andau equations in the theory of superconductivity},\
  }\href@noop {} {\bibfield  {journal} {\bibinfo  {journal} {Sov. Phys. - JETP
  (Engl. Transl.)}\ }\textbf {\bibinfo {volume} {9}},\ \bibinfo {pages} {1364}
  (\bibinfo {year} {1959})}\BibitemShut {NoStop}%
\bibitem [{\citenamefont {Tinkham}(2004)}]{Tinkham2004}%
  \BibitemOpen
  \bibfield  {author} {\bibinfo {author} {\bibfnamefont {M.}~\bibnamefont
  {Tinkham}},\ }\href@noop {} {\emph {\bibinfo {title} {Introduction to
  Superconductivity}}}\ (\bibinfo  {publisher} {Dover Publications},\ \bibinfo
  {year} {2004})\BibitemShut {NoStop}%
\bibitem [{\citenamefont {Kim}\ \emph {et~al.}(2018)\citenamefont {Kim},
  \citenamefont {Tanatar},\ and\ \citenamefont {Prozorov}}]{Kim2018rsi}%
  \BibitemOpen
  \bibfield  {author} {\bibinfo {author} {\bibfnamefont {H.}~\bibnamefont
  {Kim}}, \bibinfo {author} {\bibfnamefont {M.~A.}\ \bibnamefont {Tanatar}},\
  and\ \bibinfo {author} {\bibfnamefont {R.}~\bibnamefont {Prozorov}},\
  }\bibfield  {title} {\bibinfo {title} {Tunnel diode resonator for precision
  magnetic susceptibility measurements in a mk temperature range and large dc
  magnetic fields},\ }\href {https://doi.org/10.1063/1.5048008} {\bibfield
  {journal} {\bibinfo  {journal} {Rev. Sci. Instrum.}\ }\textbf {\bibinfo
  {volume} {89}},\ \bibinfo {pages} {094704} (\bibinfo {year}
  {2018})}\BibitemShut {NoStop}%
\bibitem [{\citenamefont {Degrift}(1975)}]{Degrift1975}%
  \BibitemOpen
  \bibfield  {author} {\bibinfo {author} {\bibfnamefont {C.~T.~V.}\
  \bibnamefont {Degrift}},\ }\bibfield  {title} {\bibinfo {title} {Tunnel diode
  oscillator for 0.001 ppm measurements at low temperatures},\ }\href
  {https://doi.org/10.1063/1.1134272} {\bibfield  {journal} {\bibinfo
  {journal} {Rev. Sci. Instr.}\ }\textbf {\bibinfo {volume} {46}},\ \bibinfo
  {pages} {599} (\bibinfo {year} {1975})}\BibitemShut {NoStop}%
\bibitem [{\citenamefont {Prozorov}\ and\ \citenamefont
  {Giannetta}(2006)}]{Prozorov2006}%
  \BibitemOpen
  \bibfield  {author} {\bibinfo {author} {\bibfnamefont {R.}~\bibnamefont
  {Prozorov}}\ and\ \bibinfo {author} {\bibfnamefont {R.~W.}\ \bibnamefont
  {Giannetta}},\ }\bibfield  {title} {\bibinfo {title} {Magnetic penetration
  depth in unconventional superconductors},\ }\href
  {http://stacks.iop.org/0953-2048/19/i=8/a=R01} {\bibfield  {journal}
  {\bibinfo  {journal} {Supercond. Sci. Techn.}\ }\textbf {\bibinfo {volume}
  {19}},\ \bibinfo {pages} {R41} (\bibinfo {year} {2006})}\BibitemShut
  {NoStop}%
\bibitem [{\citenamefont {Prozorov}\ and\ \citenamefont
  {Kogan}(2011)}]{Prozorov2011}%
  \BibitemOpen
  \bibfield  {author} {\bibinfo {author} {\bibfnamefont {R.}~\bibnamefont
  {Prozorov}}\ and\ \bibinfo {author} {\bibfnamefont {V.~G.}\ \bibnamefont
  {Kogan}},\ }\bibfield  {title} {\bibinfo {title} {London penetration depth in
  iron-based superconductors},\ }\href
  {https://doi.org/10.1088/0034-4885/74/12/124505} {\bibfield  {journal}
  {\bibinfo  {journal} {Rep. Progr. Phys.}\ }\textbf {\bibinfo {volume} {74}},\
  \bibinfo {pages} {124505} (\bibinfo {year} {2011})}\BibitemShut {NoStop}%
\bibitem [{\citenamefont {Vannette}\ \emph {et~al.}(2008)\citenamefont
  {Vannette}, \citenamefont {Sefat}, \citenamefont {Jia}, \citenamefont {Law},
  \citenamefont {Lapertot}, \citenamefont {Bud'ko}, \citenamefont {Canfield},
  \citenamefont {Schmalian},\ and\ \citenamefont {Prozorov}}]{Vannette2008}%
  \BibitemOpen
  \bibfield  {author} {\bibinfo {author} {\bibfnamefont {M.}~\bibnamefont
  {Vannette}}, \bibinfo {author} {\bibfnamefont {A.}~\bibnamefont {Sefat}},
  \bibinfo {author} {\bibfnamefont {S.}~\bibnamefont {Jia}}, \bibinfo {author}
  {\bibfnamefont {S.}~\bibnamefont {Law}}, \bibinfo {author} {\bibfnamefont
  {G.}~\bibnamefont {Lapertot}}, \bibinfo {author} {\bibfnamefont
  {S.}~\bibnamefont {Bud'ko}}, \bibinfo {author} {\bibfnamefont
  {P.}~\bibnamefont {Canfield}}, \bibinfo {author} {\bibfnamefont
  {J.}~\bibnamefont {Schmalian}},\ and\ \bibinfo {author} {\bibfnamefont
  {R.}~\bibnamefont {Prozorov}},\ }\bibfield  {title} {\bibinfo {title}
  {Precise measurements of radio-frequency magnetic susceptibility in
  ferromagnetic and antiferromagnetic materials},\ }\href
  {https://doi.org/10.1016/j.jmmm.2007.06.018} {\bibfield  {journal} {\bibinfo
  {journal} {J. Magn. Magn. Mater.}\ }\textbf {\bibinfo {volume} {320}},\
  \bibinfo {pages} {354 } (\bibinfo {year} {2008})}\BibitemShut {NoStop}%
\bibitem [{\citenamefont {Hardy}\ \emph {et~al.}(1993)\citenamefont {Hardy},
  \citenamefont {Bonn}, \citenamefont {Morgan}, \citenamefont {Liang},\ and\
  \citenamefont {Zhang}}]{Hardy1993}%
  \BibitemOpen
  \bibfield  {author} {\bibinfo {author} {\bibfnamefont {W.~N.}\ \bibnamefont
  {Hardy}}, \bibinfo {author} {\bibfnamefont {D.~A.}\ \bibnamefont {Bonn}},
  \bibinfo {author} {\bibfnamefont {D.~C.}\ \bibnamefont {Morgan}}, \bibinfo
  {author} {\bibfnamefont {R.}~\bibnamefont {Liang}},\ and\ \bibinfo {author}
  {\bibfnamefont {K.}~\bibnamefont {Zhang}},\ }\bibfield  {title} {\bibinfo
  {title} {Precision measurements of the temperature dependence of $\lambda$ in
  {YBa}$_{2}${Cu}$_{3}${O}$_{6.95}$: Strong evidence for nodes in the gap
  function},\ }\href {https://doi.org/10.1103/PhysRevLett.70.3999} {\bibfield
  {journal} {\bibinfo  {journal} {Phys. Rev. Lett.}\ }\textbf {\bibinfo
  {volume} {70}},\ \bibinfo {pages} {3999} (\bibinfo {year}
  {1993})}\BibitemShut {NoStop}%
\bibitem [{\citenamefont {Hirschfeld}\ and\ \citenamefont
  {Goldenfeld}(1993)}]{Hirschfeld1993}%
  \BibitemOpen
  \bibfield  {author} {\bibinfo {author} {\bibfnamefont {P.~J.}\ \bibnamefont
  {Hirschfeld}}\ and\ \bibinfo {author} {\bibfnamefont {N.}~\bibnamefont
  {Goldenfeld}},\ }\bibfield  {title} {\bibinfo {title} {Effect of strong
  scattering on the low-temperature penetration depth of a \textit{d} -wave
  superconductor},\ }\href {https://doi.org/10.1103/PhysRevB.48.4219}
  {\bibfield  {journal} {\bibinfo  {journal} {Phys. Rev. B}\ }\textbf {\bibinfo
  {volume} {48}},\ \bibinfo {pages} {4219} (\bibinfo {year}
  {1993})}\BibitemShut {NoStop}%
\bibitem [{\citenamefont {Cho}\ \emph {et~al.}(2022)\citenamefont {Cho},
  \citenamefont {Ko\ifmmode~\acute{n}\else \'{n}\fi{}czykowski}, \citenamefont
  {Teknowijoyo}, \citenamefont {Ghimire}, \citenamefont {Tanatar},
  \citenamefont {Mishra},\ and\ \citenamefont {Prozorov}}]{Cho2022}%
  \BibitemOpen
  \bibfield  {author} {\bibinfo {author} {\bibfnamefont {K.}~\bibnamefont
  {Cho}}, \bibinfo {author} {\bibfnamefont {M.}~\bibnamefont
  {Ko\ifmmode~\acute{n}\else \'{n}\fi{}czykowski}}, \bibinfo {author}
  {\bibfnamefont {S.}~\bibnamefont {Teknowijoyo}}, \bibinfo {author}
  {\bibfnamefont {S.}~\bibnamefont {Ghimire}}, \bibinfo {author} {\bibfnamefont
  {M.~A.}\ \bibnamefont {Tanatar}}, \bibinfo {author} {\bibfnamefont
  {V.}~\bibnamefont {Mishra}},\ and\ \bibinfo {author} {\bibfnamefont
  {R.}~\bibnamefont {Prozorov}},\ }\bibfield  {title} {\bibinfo {title}
  {Intermediate scattering potential strength in electron-irradiated
  {YBa}$_{2}${Cu}$_{3}${O}$_{7-\delta}$ from {L}ondon penetration depth
  measurements},\ }\href {https://doi.org/10.1103/PhysRevB.105.014514}
  {\bibfield  {journal} {\bibinfo  {journal} {Phys. Rev. B}\ }\textbf {\bibinfo
  {volume} {105}},\ \bibinfo {pages} {014514} (\bibinfo {year}
  {2022})}\BibitemShut {NoStop}%
\bibitem [{\citenamefont {Ormeno}\ \emph {et~al.}(2002)\citenamefont {Ormeno},
  \citenamefont {Sibley}, \citenamefont {Gough}, \citenamefont {Sebastian},\
  and\ \citenamefont {Fisher}}]{Ormeno2002}%
  \BibitemOpen
  \bibfield  {author} {\bibinfo {author} {\bibfnamefont {R.~J.}\ \bibnamefont
  {Ormeno}}, \bibinfo {author} {\bibfnamefont {A.}~\bibnamefont {Sibley}},
  \bibinfo {author} {\bibfnamefont {C.~E.}\ \bibnamefont {Gough}}, \bibinfo
  {author} {\bibfnamefont {S.}~\bibnamefont {Sebastian}},\ and\ \bibinfo
  {author} {\bibfnamefont {I.~R.}\ \bibnamefont {Fisher}},\ }\bibfield  {title}
  {\bibinfo {title} {Microwave conductivity and penetration depth in the heavy
  fermion superconductor {CeCoIn}$_{5}$},\ }\href
  {https://doi.org/10.1103/PhysRevLett.88.047005} {\bibfield  {journal}
  {\bibinfo  {journal} {Phys. Rev. Lett.}\ }\textbf {\bibinfo {volume} {88}},\
  \bibinfo {pages} {047005} (\bibinfo {year} {2002})}\BibitemShut {NoStop}%
\bibitem [{\citenamefont {Truncik}\ \emph {et~al.}(2013)\citenamefont
  {Truncik}, \citenamefont {Huttema}, \citenamefont {Turner}, \citenamefont
  {{\"O}zcan}, \citenamefont {Murphy}, \citenamefont {Carri{\`e}re},
  \citenamefont {Thewalt}, \citenamefont {Morse}, \citenamefont {Koenig},
  \citenamefont {Sarrao},\ and\ \citenamefont {Broun}}]{Truncik2013}%
  \BibitemOpen
  \bibfield  {author} {\bibinfo {author} {\bibfnamefont {C.~J.~S.}\
  \bibnamefont {Truncik}}, \bibinfo {author} {\bibfnamefont {W.~A.}\
  \bibnamefont {Huttema}}, \bibinfo {author} {\bibfnamefont {P.~J.}\
  \bibnamefont {Turner}}, \bibinfo {author} {\bibfnamefont {S.}~\bibnamefont
  {{\"O}zcan}}, \bibinfo {author} {\bibfnamefont {N.~C.}\ \bibnamefont
  {Murphy}}, \bibinfo {author} {\bibfnamefont {P.~R.}\ \bibnamefont
  {Carri{\`e}re}}, \bibinfo {author} {\bibfnamefont {E.}~\bibnamefont
  {Thewalt}}, \bibinfo {author} {\bibfnamefont {K.~J.}\ \bibnamefont {Morse}},
  \bibinfo {author} {\bibfnamefont {A.~J.}\ \bibnamefont {Koenig}}, \bibinfo
  {author} {\bibfnamefont {J.~L.}\ \bibnamefont {Sarrao}},\ and\ \bibinfo
  {author} {\bibfnamefont {D.~M.}\ \bibnamefont {Broun}},\ }\bibfield  {title}
  {\bibinfo {title} {Nodal quasiparticle dynamics in the heavy fermion
  superconductor {CeCoIn}$_5$ revealed by precision microwave spectroscopy},\
  }\href {https://doi.org/10.1038/ncomms3477} {\bibfield  {journal} {\bibinfo
  {journal} {Nature Comm.}\ }\textbf {\bibinfo {volume} {4}},\ \bibinfo {pages}
  {2477} (\bibinfo {year} {2013})}\BibitemShut {NoStop}%
\bibitem [{\citenamefont {Hashimoto}\ \emph {et~al.}(2013)\citenamefont
  {Hashimoto}, \citenamefont {Mizukami}, \citenamefont {Katsumata},
  \citenamefont {Shishido}, \citenamefont {Yamashita}, \citenamefont {Ikeda},
  \citenamefont {Matsuda}, \citenamefont {Schlueter}, \citenamefont {Fletcher},
  \citenamefont {Carrington}, \citenamefont {Gnida}, \citenamefont
  {Kaczorowski},\ and\ \citenamefont {Shibauchi}}]{Hashimoto2013}%
  \BibitemOpen
  \bibfield  {author} {\bibinfo {author} {\bibfnamefont {K.}~\bibnamefont
  {Hashimoto}}, \bibinfo {author} {\bibfnamefont {Y.}~\bibnamefont {Mizukami}},
  \bibinfo {author} {\bibfnamefont {R.}~\bibnamefont {Katsumata}}, \bibinfo
  {author} {\bibfnamefont {H.}~\bibnamefont {Shishido}}, \bibinfo {author}
  {\bibfnamefont {M.}~\bibnamefont {Yamashita}}, \bibinfo {author}
  {\bibfnamefont {H.}~\bibnamefont {Ikeda}}, \bibinfo {author} {\bibfnamefont
  {Y.}~\bibnamefont {Matsuda}}, \bibinfo {author} {\bibfnamefont {J.~A.}\
  \bibnamefont {Schlueter}}, \bibinfo {author} {\bibfnamefont {J.~D.}\
  \bibnamefont {Fletcher}}, \bibinfo {author} {\bibfnamefont {A.}~\bibnamefont
  {Carrington}}, \bibinfo {author} {\bibfnamefont {D.}~\bibnamefont {Gnida}},
  \bibinfo {author} {\bibfnamefont {D.}~\bibnamefont {Kaczorowski}},\ and\
  \bibinfo {author} {\bibfnamefont {T.}~\bibnamefont {Shibauchi}},\ }\bibfield
  {title} {\bibinfo {title} {Anomalous superfluid density in quantum critical
  superconductors},\ }\href {https://doi.org/10.1073/pnas.1221976110}
  {\bibfield  {journal} {\bibinfo  {journal} {Proc. Nat. Acad. Sci.}\ }\textbf
  {\bibinfo {volume} {110}},\ \bibinfo {pages} {3293} (\bibinfo {year}
  {2013})}\BibitemShut {NoStop}%
\bibitem [{\citenamefont {Paglione}\ \emph {et~al.}(2003)\citenamefont
  {Paglione}, \citenamefont {Tanatar}, \citenamefont {Hawthorn}, \citenamefont
  {Boaknin}, \citenamefont {Hill}, \citenamefont {Ronning}, \citenamefont
  {Sutherland}, \citenamefont {Taillefer}, \citenamefont {Petrovic},\ and\
  \citenamefont {Canfield}}]{Paglione2003}%
  \BibitemOpen
  \bibfield  {author} {\bibinfo {author} {\bibfnamefont {J.}~\bibnamefont
  {Paglione}}, \bibinfo {author} {\bibfnamefont {M.~A.}\ \bibnamefont
  {Tanatar}}, \bibinfo {author} {\bibfnamefont {D.~G.}\ \bibnamefont
  {Hawthorn}}, \bibinfo {author} {\bibfnamefont {E.}~\bibnamefont {Boaknin}},
  \bibinfo {author} {\bibfnamefont {R.~W.}\ \bibnamefont {Hill}}, \bibinfo
  {author} {\bibfnamefont {F.}~\bibnamefont {Ronning}}, \bibinfo {author}
  {\bibfnamefont {M.}~\bibnamefont {Sutherland}}, \bibinfo {author}
  {\bibfnamefont {L.}~\bibnamefont {Taillefer}}, \bibinfo {author}
  {\bibfnamefont {C.}~\bibnamefont {Petrovic}},\ and\ \bibinfo {author}
  {\bibfnamefont {P.~C.}\ \bibnamefont {Canfield}},\ }\bibfield  {title}
  {\bibinfo {title} {Field-induced quantum critical point in {CeCoIN}$_{5}$},\
  }\href {https://doi.org/10.1103/PhysRevLett.91.246405} {\bibfield  {journal}
  {\bibinfo  {journal} {Phys. Rev. Lett.}\ }\textbf {\bibinfo {volume} {91}},\
  \bibinfo {pages} {246405} (\bibinfo {year} {2003})}\BibitemShut {NoStop}%
\bibitem [{\citenamefont {Brandt}(1991)}]{Brandt1991}%
  \BibitemOpen
  \bibfield  {author} {\bibinfo {author} {\bibfnamefont {E.~H.}\ \bibnamefont
  {Brandt}},\ }\bibfield  {title} {\bibinfo {title} {Penetration of magnetic ac
  fields into type-{II} superconductors},\ }\href
  {https://doi.org/10.1103/PhysRevLett.67.2219} {\bibfield  {journal} {\bibinfo
   {journal} {Phys. Rev. Lett.}\ }\textbf {\bibinfo {volume} {67}},\ \bibinfo
  {pages} {2219} (\bibinfo {year} {1991})}\BibitemShut {NoStop}%
\bibitem [{\citenamefont {Chia}\ \emph {et~al.}(2003)\citenamefont {Chia},
  \citenamefont {Van~Harlingen}, \citenamefont {Salamon}, \citenamefont
  {Yanoff}, \citenamefont {Bonalde},\ and\ \citenamefont {Sarrao}}]{Chia2003}%
  \BibitemOpen
  \bibfield  {author} {\bibinfo {author} {\bibfnamefont {E.~E.~M.}\
  \bibnamefont {Chia}}, \bibinfo {author} {\bibfnamefont {D.~J.}\ \bibnamefont
  {Van~Harlingen}}, \bibinfo {author} {\bibfnamefont {M.~B.}\ \bibnamefont
  {Salamon}}, \bibinfo {author} {\bibfnamefont {B.~D.}\ \bibnamefont {Yanoff}},
  \bibinfo {author} {\bibfnamefont {I.}~\bibnamefont {Bonalde}},\ and\ \bibinfo
  {author} {\bibfnamefont {J.~L.}\ \bibnamefont {Sarrao}},\ }\bibfield  {title}
  {\bibinfo {title} {Nonlocality and strong coupling in the heavy fermion
  superconductor {CeCoIn}$_{5}:$ a penetration depth study},\ }\href
  {https://doi.org/10.1103/PhysRevB.67.014527} {\bibfield  {journal} {\bibinfo
  {journal} {Phys. Rev. B}\ }\textbf {\bibinfo {volume} {67}},\ \bibinfo
  {pages} {014527} (\bibinfo {year} {2003})}\BibitemShut {NoStop}%
\bibitem [{\citenamefont {Labusch}(1969)}]{Labusch1969}%
  \BibitemOpen
  \bibfield  {author} {\bibinfo {author} {\bibfnamefont {R.}~\bibnamefont
  {Labusch}},\ }\bibfield  {title} {\bibinfo {title} {Calculation of the
  critical field gradient in type-{II} superconductors},\ }\href
  {https://www.osti.gov/biblio/4133649} {\bibfield  {journal} {\bibinfo
  {journal} {Cryst. Lattice Defects}\ }\textbf {\bibinfo {volume} {1}},\
  \bibinfo {pages} {1} (\bibinfo {year} {1969})}\BibitemShut {NoStop}%
\end{thebibliography}
\end{document}